\documentclass[aps,twocolumn,showpacs,preprintnumbers,amsmath,amssymb]{revtex4}
\usepackage{graphicx}

\usepackage{dcolumn}

\usepackage{bm}
\newcommand{\sel}{s_e}
\newcommand{\spi}{s_{\pi}}
\newcommand{\thpi}{\theta_\pi}
\newcommand{\thel}{\theta_e}
\newcommand{\co}{\; ,}
\newcommand{\po}{\; .}
\newcommand{\nn}{\nonumber \\}
\begin{document}
\vskip 1cm

\title{\bf High statistics measurement of $K_{e4}$ decay properties}
\author{S. Pislak$^{7,6}$, R. Appel$^{6,3}$, G.S. Atoyan$^4$, 
B. Bassalleck$^2$, D.R. Bergman$^6$\cite{DB}, N. Cheung$^3$,  \\
S. Dhawan$^6$, H. Do$^6$, J. Egger$^5$, S. Eilerts$^2$\cite{SE},
H. Fischer$^2$\cite{HF}, W. Herold$^5$, \\
V.V. Issakov$^4$, H. Kaspar$^{5,6}$, D.E. Kraus$^3$, D.M. Lazarus$^1$,
P. Lichard$^3$, J. Lowe$^2$, \\
J. Lozano$^6$\cite{JL}, H. Ma$^1$, W. Majid$^6$\cite{WMa},
 A.A. Poblaguev$^4$, P. Rehak$^1$, A. Sher$^3$\cite{AS}, \\
Aleksey Sher$^7$, J.A. Thompson$^3$, P. Tru\"ol$^{7,6}$, and 
M.E. Zeller$^6$   \\
}

\affiliation{
$^1$Brookhaven National Laboratory, Upton L. I., NY 11973, USA\\ 
$^2$Department of Physics and Astronomy, 
University of New Mexico, Albuquerque, NM 87131, USA\\
$^3$Department of Physics and Astronomy, University of Pittsburgh,
Pittsburgh, PA 15260, USA \\ 
$^4$Institute for Nuclear Research of Russian Academy of Sciences, 
Moscow 117 312, Russia \\
$^5$Paul Scherrer Institut, CH-5232 Villigen, Switzerland\\ 
$^6$Physics Department, Yale University, New Haven, CT 06511, USA\\
$^7$Physik-Institut, Universit\"at Z\"urich, CH-8057 Z\"urich, Switzerland}

\date{\today}

\vskip 0.5cm
\begin{abstract}
We report experimental details and results of 
a new measurement of the decay 
$K^+\rightarrow\pi^+\pi^-e^+\nu_e$ ($K_{e4}$). A sample of more than 400,000 
$K_{e4}$ events with low background has been collected by Experiment 865
at the Brookhaven Alternate Gradient Synchrotron. From these data,
the branching ratio $(4.11\pm0.01\pm0.11)\cdot 10^{-5}$ and the
$\pi\pi$ invariant mass dependence 
of the form factors $F$, $G$, and $H$ of the 
weak hadronic current as well as the phase shift difference
$\delta^0_0-\delta^1_1$ for $\pi\pi$-scattering were extracted.
Using constraints based on analyticity and chiral symmetry, a new value 
with considerably improved accuracy for the
$s$-wave $\pi\pi$-scattering length $a^0_0$ has been obtained also: 
$a_0^0=0.216\pm0.013\,({\rm stat.})\pm0.002\,({\rm syst.})
\pm0.002\,({\rm theor.})$. 
\end{abstract}
\pacs{PACS numbers: 13.20.-v, 13.20.Eb, 13.75.Lb}
\maketitle

\section{Introduction}
\label{sec:introduction}

Among the long list of possible charged kaon decays the rare
$K_{e4}$ decay branch $(K^\pm\rightarrow \pi^+\pi^-e^\pm\nu_e
(\bar{\nu}_e))$ has received particular attention,
because it was recognized~\cite{shabalin63}, almost coincident with the
observation of the first event for this decay 40 years ago~\cite{koller62},
that it could provide important information
on the structure of the weak hadronic currents and also on $\pi\pi$ scattering
at low energies. The final state interaction of the two pions was expected
to manifest itself in an angular correlation between the decay products,
namely an asymmetry of the lepton distribution with respect to the plane
formed by the two pion momenta. This asymmetry is directly related
to the difference between the $s$- and $p$-wave scattering phase.
What made this four-body semileptonic decay attractive despite its
low branching ratio, which was then predicted to be of order 
$10^{-5}$~\cite{okun60}, is that the 
two pions are the only hadrons in the final state. For all other
reactions used to study the $\pi\pi$ interaction,
e.g. $\pi^- p\rightarrow \pi^-\pi^+ n$, there is at least one other
hadron present in the final state.
Thus experimental studies of the $K_{e4}$-decay were seen as the 
cleanest method 
to determine the isospin zero, angular momentum zero scattering length 
$a_0^0$. Since early 
experiments~\cite{birge65,ely69,schweinberger71,bourquin71,beier72} observed
only a few hundred events each, 
it was not until 1977, when the  
Geneva-Saclay experiment~\cite{rosselet77} gathered about 30,000 events,
that a measurement was made of this quantity to 20\% accuracy.

Since then no new data became available until Experiment 865 at the
Brookhaven Alternating Gradient Synchrotron collected 400,000 $K_{e4}$
events. We report here the details of the analysis of these data, some of
which have been communicated earlier~\cite{pislak01}. 
A promising alternative  way to study 
$\pi\pi$-interactions through a measurement of the lifetime of the
$\pi\pi$-atom is followed in the DIRAC experiment at 
CERN~\cite{dirac}, which has not yet yielded a definitive result.

The theoretical analysis of $\pi\pi$-interactions at low energies
is intimately linked to the development of chiral quantum chromodynamics
perturbation theory (ChPT)~\cite{ChPT,leutwyler01,honnef02}. 
In this approach, the fact that standard QCD perturbation
theory is not directly applicable at low energies because
the strong coupling becomes large, is circumvented through 
a systematic expansion of the observables in terms of 
external momenta and of light quark masses. The 
spontaneous breakdown of the underlying chiral symmetry is
associated with the quark-antiquark vacuum expectation value,
the so-called quark condensate $\langle0|\overline{q}q|0\rangle$.
It is normally assumed to be of natural size, or equivalently
that the Gell-Mann--Oakes--Renner formula~\cite{gellmann68} for the pion mass 
\begin{equation}  m_\pi^2\simeq \frac{1}{F_\pi^2}(m_u+m_d)
\langle0|\overline{q}q|0\rangle \label{equ:oakes} \end{equation}
has only small corrections. Here $F_{\pi}\simeq93$~MeV is the pion decay 
constant. This assumption does not have to be
made, as the authors of a less restrictive
version of chiral perturbation theory (GChPT)~\cite{GChPT,knecht95}
pointed out.
The measurement of the
$\pi\pi$ threshold parameters has been advocated as one of the areas
where a significant difference between the two approaches could be observed.
ChPT, however, makes firm predictions for 
the scattering length. The tree level calculation
($\mathcal{O}(p^2)$~\cite{weinberg66}) yields
$a_0^0=0.156$ (in this paper we use units of $m_\pi^{-1}$
for the scattering length). The one-loop 
($\mathcal{O}(p^4)$, $a_0^0=0.201\pm0.01$~\cite{gasser83}) and the 
two loop calculation (($\mathcal{O}(p^6)$, $a_0^0=0.217$~\cite{bijnens96}) 
show  satisfactory 
convergence. The most recent calculation~\cite{colangelo00,colangelo01b} 
matches the known chiral perturbation theory representation 
of the $\pi\pi$ scattering amplitude to two loops~\cite{bijnens96} 
with the dispersive representation that follows from
the Roy equations~\cite{roy71,ananthanarayan00},
resulting in the prediction $a_0^0=0.220 \pm 0.005$. 
The high precision of this prediction has to
be contrasted with the experimental value 
$a_0^0=(0.26\pm0.05)$ extracted
from the Geneva-Saclay experiment~\cite{rosselet77}
using the Roy equations and some peripheral
$\pi N\rightarrow \pi\pi N$ data~\cite{nagels79}.

The form factors appearing in the
weak hadronic current in the $K_{e4}$ decay matrix element,
have also been extensively used for the
determination of the parameters of the ChPT 
Hamiltonian~\cite{talavera00,talavera01}.
This program would clearly benefit from lower experimental
uncertainties.

\section{Theoretical background for the analysis of $K_{e4}$ decay}
\label{sec:theory}
\subsection{Kinematics}
\label{sec:theory_kinematics}

\begin{figure}[htb]
\includegraphics[width=70mm]{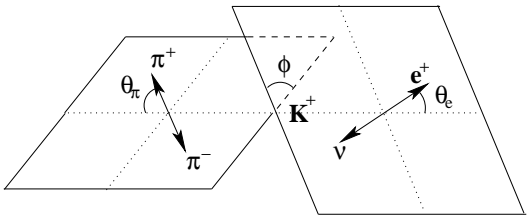}\centering
\caption[Kinematics of the $K_{e4}$ decay]
{Kinematic quantities used in the analysis of $K_{e4}$ decay.
\label{fig:kinematics}}
\end{figure}

The decay
\begin{equation}
\label{equ:decay}
K^+(p)\rightarrow \pi^+(p_1)\pi^-(p_2)e^+(p_e)\nu_e(p_{\nu})
\end{equation}
can most conveniently be treated~\cite{ke494} 
by using three reference frames,
as illustrated in Fig.~\ref{fig:kinematics}:
(1) the $K^+$ rest system ($\Sigma_K$), (2) the $\pi^+\pi^-$ rest system
($\Sigma_{\pi\pi}$) and (3) the $e^+\nu_e$ rest system ($\Sigma_{e\nu}$). 
The kinematics of the $K_{e4}$ decay are then fully described by five 
variables, introduced by Cabibbo and Maksymowicz~\cite{cabibbo65}:
\begin{enumerate}\setlength{\itemsep}{-3pt}
\item $\spi=M_{\pi\pi}^2$, the invariant mass squared of the dipion;
\item $\sel=M_{e\nu}^2$, the invariant mass squared of the dilepton;
\item $\thpi$, the angle of the $\pi^+$ in $\Sigma_{\pi\pi}$ with respect 
to the direction of flight of the dipion in $\Sigma_K$;
\item $\thel$, the angle of the $e^+$ in $\Sigma_{e\nu}$ with respect to 
the direction of flight of the dilepton in $\Sigma_K$;
\item $\phi$, the angle between the plane formed by the two pions and the 
corresponding plane formed by the two leptons.
\end{enumerate}

It is useful for the following discussion to introduce the combinations
$P$, $Q$ and $L$
of the momentum four vectors $p_1$, $p_2$, $p_e$ and $p_\nu$  
defined in Eq.~(\ref{equ:decay}) and two scalar products derived from them
\begin{eqnarray}
P=p_1+p_2\;, \; \; Q=p_1-p_2\;,\; \; L=p_e+p_\nu\;,\\
Q^2=4m_{\pi}^2-\spi\;,\;\;
P\cdot L = \frac{1}{2}(m_K^2-\spi-\sel)\;,\\ 
X=[(P\cdot L)^2-\spi\sel]^{1/2}\;,\;\;
\sigma_\pi = (1-4m_\pi^2/s_\pi)^{1/2}\;.  
\end{eqnarray}

\subsection{Matrix element}
\label{sec:theory_matrixelement}
The matrix element is written as
\begin{equation}
\label{equ:matrixelement}
M=\frac{G_F}{\sqrt{2}} V^*_{us} \overline{u}(p_{\nu})\gamma_{\mu}
(1-\gamma_5)v(p_e)(V^{\mu}-A^{\mu}) \po
\end{equation}
The vector current $V^{\mu}$ and the axial vector current $A^{\mu}$ 
have to be Lorentz invariant four-vectors: 
\begin{eqnarray}
A^{\mu}&=&\frac{1}{m_K}\left(F P^{\mu}+G Q^{\mu} + R L^{\mu}\right) \co \nn
V^{\mu}&=&\frac{H}{m_K^3}\epsilon^{\mu\nu\rho\sigma}
          L_{\nu}P_{\rho}Q_{\sigma}\po 
\end{eqnarray}
The kaon mass $m_K$ was inserted to make the form factors $F$, $G$, $R$ 
and $H$ dimensionless complex functions of $p_1\cdot p_2$, $p_1 \cdot p$ 
and $p_2 \cdot p$ or equivalent of $\spi$, $\sel$ and $\thpi$.

\subsection{Decay rate}
\label{sec:theory_decayrate}

The decay rate following from the
matrix element given in Eq.~\ref{equ:matrixelement} and neglecting
terms proportional to $m_2^e/s_e$ is
given by~\cite{pais67}
\begin{widetext}
\begin{eqnarray}\label{equ:decayrate}
d\Gamma_5& =& \frac{G_F^2 V^2_{us}}{2^{12}\pi^6m_K^5}X
\sigma_\pi J_5(\spi,\sel,\thpi,\thel,\phi)
d\spi d\sel d(\cos\thpi) d(\cos\thel) d\phi \co\\
\label{equ:i}
J_5&=&I_1+I_2\cos{2\theta_e} + I_3\sin^2{\theta_e}\cos{2\phi} 
  + I_4 \sin{2\theta_e}\cos{\phi} + I_5 \sin{\theta_e}\cos{\phi} \nn
 && + I_6 \cos{\theta_e} + I_7\sin{\theta_e}\sin{\phi}  
   + I_8 \sin{2\theta_e}\sin{\phi} + I_9\sin^2{\theta_e}\sin^2{\phi} \co 
\end{eqnarray}
\end{widetext}

\begin{widetext}
Again neglecting terms proportional to $m_e^2/s_e$ the 
functions $I_i$ are given by
\begin{eqnarray}
\label{equ:ii}
I_1 &= & \frac{1}{8}\left\{ 2 |F_1|^2  + 3\left(|F_2|^2 +
|F_3|^2 \right) \sin^2 \thpi   \right\}\co
\hspace*{3mm} 
I_2 = - \frac{1}{8} \left\{ 2|F_1|^2 - 
\left( |F_2|^2 + |F_3|^2 \right)  \sin^2 \thpi  \right\}
\co \\
I_3 &=& - \frac{1}{4}  \left\{ |F_2|^2 
- |F_3|^2 \right\}\sin^2 \thpi \co\hspace*{3mm}
 I_4 = \frac{1}{2}\mbox{ Re} (F_1^* F_2)  \sin \thpi
\co \hspace*{3mm}I_5 = - \mbox{ Re} (F_1^* F_3)   \sin \thpi
\co \nn
I_6 &=&- \mbox{ Re} (F_2^* F_3)\sin^2 \thpi  
\co \hspace*{3mm}
I_7 = -  \mbox{ Im} (F_1^* F_2)  \sin \thpi
\co \hspace*{3mm}
I_8 = \frac{1}{2} \mbox{ Im} (F_1^* F_3) \sin \thpi
\co \hspace*{3mm}
I_9 = -\frac{1}{2} \mbox{ Im} (F_2^* F_3) \sin^2 \thpi \ .\nonumber
\end{eqnarray}
\end{widetext}
The form factors $F$, $G$, and $H$ are contained in the functions
$F_i$,
which are given by
\begin{eqnarray}
\label{equ:f}
F_1& =& X F + \sigma_\pi (P\cdot L) \cos \thpi \cdot G
\co \nn
F_2& =& \sigma_\pi(\spi\sel)^{1/2} G
\co \nn
F_3& =& \sigma_\pi X (\spi\sel)^{1/2} \frac{H}{m_K^2}
\end{eqnarray}
The contribution of the form factor $R$ is suppressed by a factor
$m_e^2/s_e$ and is therefore negligible. Consequently $R$ cannot be 
determined from $K_{e4}$ decay.

\subsection{Parametrisation of the form factors}
\label{sec:theory_parametrisation}
As noted above, the form factors $F$, $G$ and $H$ are functions of 
$\theta_{\pi}$, $s_{\pi}$ and $s_e$, and can be determined directly
from a fit to the experimental data for sufficiently small bins
of these kinematic variables. Alternatively a
parametrisation recently introduced by 
Amor\'os and Bijnens~\cite{amoros99} may be used, which is based 
on a partial wave expansion in the variable $\theta_{\pi}$:
\begin{eqnarray}
F&=&\left(f_s+f_s^\prime \,q^2+f_s^{\prime\prime}\, q^4+f_e 
(s_e/4m_\pi^2)\right)
    e^{i\delta^0_0(s_\pi)} \nn
 &+& \tilde{f}_p \,(\sigma_\pi X/4m_\pi^2)\cos\theta_\pi \,
    e^{i\delta^1_1(s_\pi)} \nn
G&=&\left(g_p + g_p^\prime \,q^2 + g_e(s_e/4m_\pi^2) \right) e^{i\delta^1_1(s_\pi)} \nn
H&=&\left(h_p+h_p^\prime \,q^2\right) e^{i\delta^1_1(s_\pi)} \co
\label{equ:amoros}
\end{eqnarray}     
where $q=[(\spi-4m_\pi^2)/4m_\pi^2]^{1/2}$ is the pion momentum in 
$\Sigma_{\pi\pi}$.
This parametrisation was constrained by theoretical models and the 
expected accuracy of the experimental data. It yields 10 new 
dimensionless form factor parameters 
$f_s$, $f_s^\prime$, 
$f_s^{\prime\prime}$, $f_e$, $\tilde{f}_p$, $g_p$, $g_p^\prime$, $g_e$,
$h_p$, and $h_p^\prime$, which do not depend on any kinematic
variables, plus two phase shifts, which can be identified using
Watson's theorem~\cite{watson52} with 
the $s$- and $p$-wave (isoscalar and isovector, respectively) 
$\pi\pi$ scattering phase shifts $\delta_0^0$ and 
$\delta_1^1$, which are still functions of $s_\pi$. 
In our analysis we will additionally assume $f_e=\tilde{f}_e=g_e=h_p^\prime=0$.
The validity of this assumption will be experimentally tested.
When Eq.~\ref{equ:amoros} is inserted into Eq.~\ref{equ:f} and then
into Eq.~\ref{equ:ii}, it can be observed that the 
phase shift difference $\delta=\delta_0^0-\delta_1^1$ enters via
$\cos\delta$ into the
terms $I_1$, $I_2$, $I_4$, $I_5$, and via $\sin\delta$ into the 
terms $I_7$ and $I_8$. 
Since $\delta< 0.3$ with $\cos\delta>0.95$ holds in $K_{e4}$-decay, and the
kinematic factors suppress the term $I_8$, only the term $I_7$
is really relevant, which appears 
in the decay rate (Eq.~\ref{equ:i} and Eq.~\ref{equ:decayrate}) 
multiplied by $\sin\phi$. $I_7$ and $I_8$ are the only odd $\phi$ terms. 
Hence, as noted by Shabalin~\cite{shabalin63}, and Pais and 
Treiman~\cite{pais67}, the asymmetry of the $\phi$
distribution is the observable that is most sensitive to 
the phase shifts. This also holds for any other
parametrisation of the form factors. 
The amplitude of the asymmetry is quite small compared
to the $\phi$ independent part, as Figures \ref{fig:phifits}
and \ref{fig:mccompkin} illustrate. This explains
why a very high statistics data sample is needed for an accurate
measurement of the phase shift difference.

\subsection{$\pi\pi$ Scattering length}
\label{sec:theory_scatt_length}

To establish a relation between the phase shift $\delta_0^0$ 
and the scattering length normally the analytical properties of the
$\pi\pi$ scattering amplitudes and crossing relations are used, which lead
to dispersion relations contained in the Roy equations~\cite{roy71}. 
Ananthanarayan~{\it et al.}~\cite{ananthanarayan00} have recently updated
earlier treatments~\cite{morgan94}, which were used in the analysis
of $\pi\pi$ scattering data, and solved these equations numerically.
Their analysis made use of a phase shift
parametrisation originally proposed by Schenk~\cite{schenk91}:
\begin{eqnarray}
\tan \delta_\ell^I &=& \sqrt{1-\frac{4 m_\pi^2}{s_\pi}}\; q^{2 \ell} 
\left\{A^I_\ell
+ B^I_\ell q^2 + C^I_\ell q^4 + D^I_\ell q^6 \right\} \nn
  &\times&\left(\frac{4m_\pi^2 - s^I_\ell}{s_\pi-s^I_\ell} \right) \,.
\label{equ:schenk}
\end{eqnarray}
The solution of the Roy equations implies that the parameters
$A^I_\ell$, $B^I_\ell$, etc. can be expressed as a function of only two
parameters or subtraction constants, which are identified as
the $I=0$ and $I=2$ $s$-wave scattering lengths
$a_0^0$ and $a_0^2$. For example, the first two coefficients
of this expression for
the $I=\ell=0$ case read as follows~\cite{coeff}  
\begin{eqnarray*}
A^0_0&=&a_0^0\;,\nn
B_0^0&=&0.2395+0.9237\Delta a_0^0-3.352\Delta a_0^2+0.2817(\Delta a_0^0)^2 \nn
&&+6.335(\Delta a_0^2)^2+6.074\Delta a_0^0\Delta a_0^2+\ldots
\;,\nn
s_0^0&= &36.83 m_\pi^2\left(1+0.2764\Delta a_0^0-0.1409\Delta a_0^2+
\ldots\right)\nn
&\simeq &(0.847)^2 \;{\rm GeV}^2\;,
\label{equ:schenknumbers}
\end{eqnarray*}
where $\Delta a_0^0\equiv a_0^0-0.220$ and $\Delta a_0^2\equiv a_0^2+0.0444$.
Although $K_{e4}$ decay allows only $I=0$ and $I=1$ contributions, the use of
the crossing relations brings in a modest dependence on the $I=2$ scattering
length. The $I=1$ phase shifts at low energies are dominated by the
$\rho$ resonance and are furthermore small in the region of interest for
$K_{e4}$.

\begin{figure}[htb]
\includegraphics[width=0.85\linewidth]{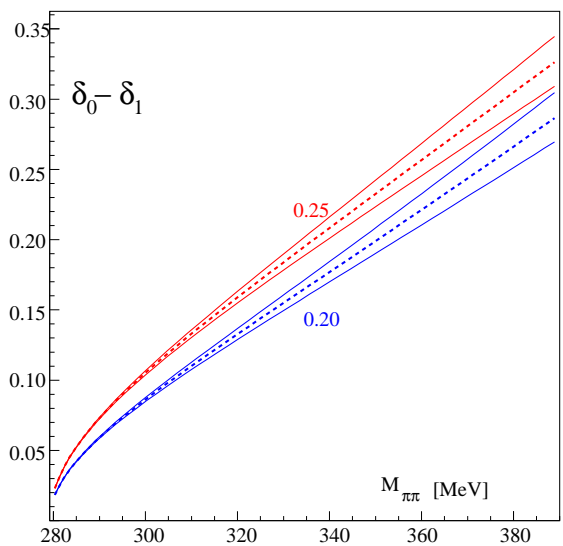}\centering
\caption[Phase shifts and scattering length]
{Predictions for the phase shift $\delta$
resulting from Eq.~\ref{equ:schenk} and Eq.~\ref{equ:universalcurve} for two
values of $a_0^0$. The three curves refer to the upper and lower limit
and the center, respectively, of the universal band in the
$(a_0^2,a^0_0)$-plane (Eq.~\ref{equ:universalcurve}).
\label{fig:universalband}}
\end{figure}

It was recognized by Morgan and Shaw~\cite{morgan69} that the possible 
values of $a_0^0$ and $a_0^2$ are restricted to a band in the 
$a_0^0-a_0^2$-plane, the so-called {\em universal band}. This band is 
defined as the area
which is allowed by $\pi\pi$ scattering data above 
0.8~GeV~\cite{hyams73,protopopescu73} and 
the Roy equations. The allowed range, estimated in the most recent 
analysis~\cite{ananthanarayan00},
is shown in Fig.~\ref{fig:ellipse}. The central curve 
of this band is given by
\begin{equation} 
a_0^2 =-0.0849+0.232\, a_0^0-0.0865\, (a_0^0)^2\; \left[\pm 0.0088\right] \;,
\label{equ:universalcurve}
\end{equation}
where the figure given in the bracket indicates the width of the band.
Figure~\ref{fig:universalband} illustrates the influence of the
{\em universal band} and  how the phase shift difference
$\delta=\delta_0^0-\delta_1^1$ depends on the scattering
length $a_0^0$.
 
It has recently been shown by 
Colangelo {\it et al.}~\cite{colangelo01a,colangelo01b}
that the width of the allowed band can be considerably reduced
to  $\left[\pm 0.0008\right]$, if
chiral symmetry constraints are imposed in 
addition. $a_0^2$ and $a_0^0$ are then related as
\begin{eqnarray} 
\Delta a_0^2 = 0.236\Delta a_0^0 -0.61(\Delta a_0^0)^2 
-9.9 (\Delta a_0^0)^3\;, 
\label{equ:universalcurvegilberto}
\end{eqnarray} 
where $\Delta a_0^2$ and $\Delta a_0^0$ have been defined above.
This band is also depicted in Fig.~\ref{fig:ellipse} with the
label CLG.

In ChPT up to order $\mathcal{O}(p^4)$ the scattering lengths
are linked to two coupling constants $\ell_3$ and $\ell_4$.
For example, $\ell_3$ determines the size of the first order correction
to the Gell-Mann--Oakes-Renner relation~(Eq.~\ref{equ:oakes})
~\cite{gellmann68}, and
is assumed to be a priori unknown in GChPT.
Colangelo~{\it et al.}~\cite{colangelo01a,colangelo01b} have argued, that
both $a_0^0$ and $a_0^2$ can be made dependent solely on $\ell_3$, if
the scalar radius of the pion is used as an additional input to give
a relation between $\ell_3$ and $\ell_4$. This also holds in GChPT, and 
Eq.~\ref{equ:universalcurvegilberto} results, when $\ell_4$ is eliminated. 
Once the scattering lengths are known experimentally, 
a constraint for $\ell_3$ and consequently for the quark condensate can 
be derived. 

\section{Experimental set up}
\label{sec:experiment}
\subsection{Apparatus}
\label{sec:apparatus}
\begin{figure}[htb]
\includegraphics[width=0.98\linewidth]{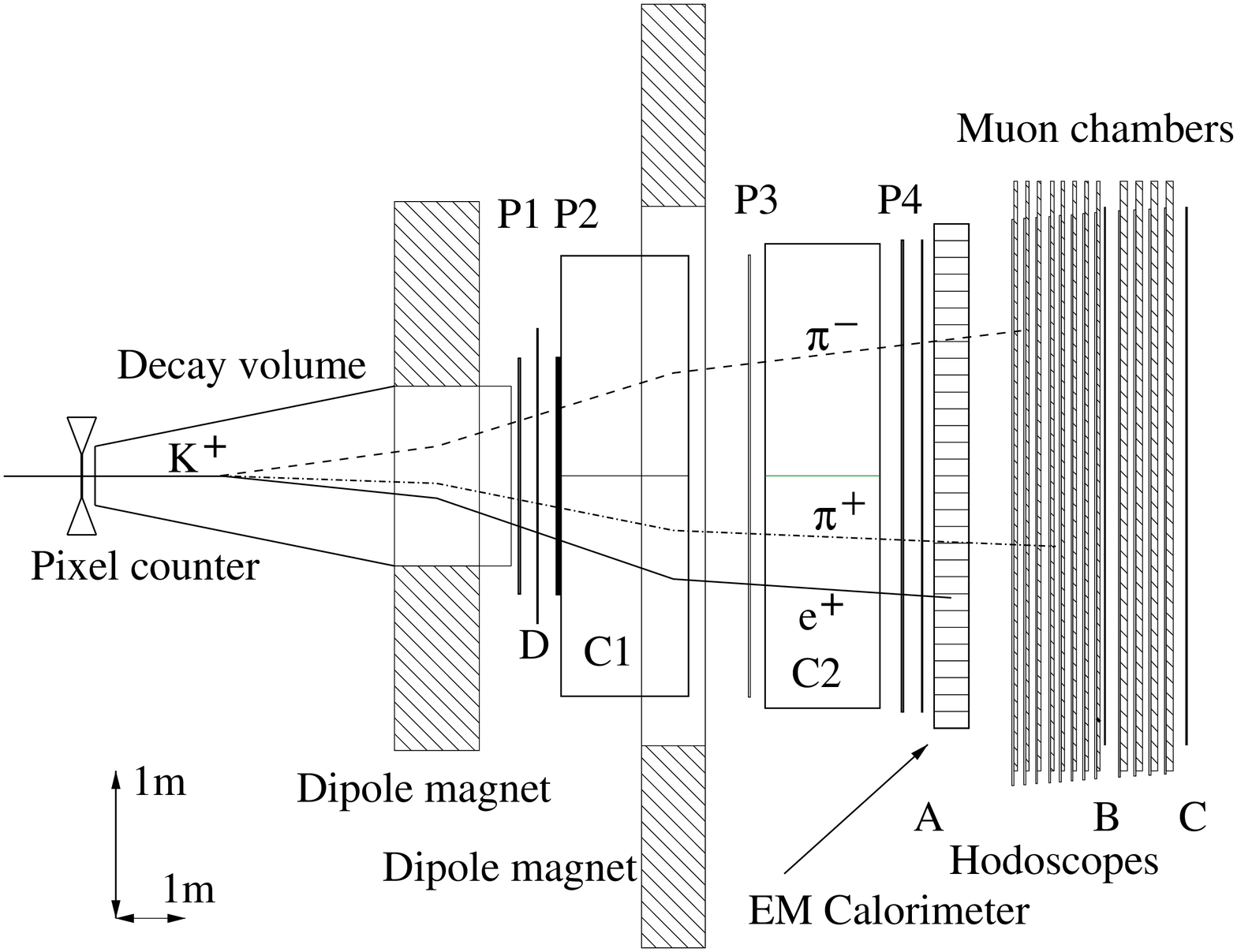}\centering
\caption[Detector.]
{Plan view of the E865 detector. A $K_{e4}$ event is superimposed.}
\label{fig:detector}
\end{figure}

The analysis outlined here is based on data recorded at the 
Brookhaven Alternate Gradient Synchrotron (AGS) in a dedicated run at
reduced beam intensity in 1997, employing the E865
detector. The apparatus, described in great detail in~\cite{appel02},
is shown in Fig.~\ref{fig:detector}.
\begin{figure*}[htb]
\includegraphics[width=0.33\linewidth]{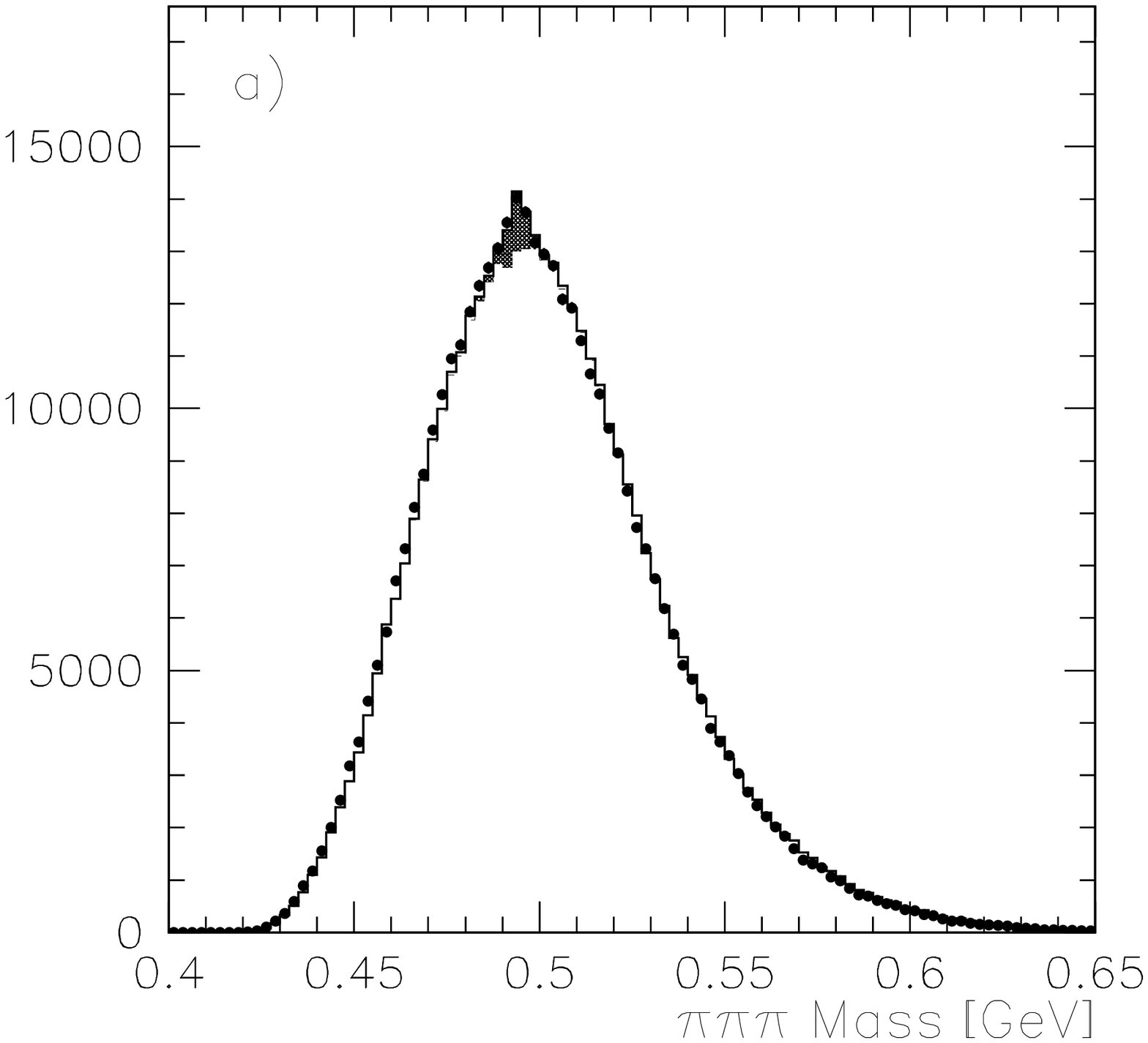}\centering
\includegraphics[width=0.33\linewidth]{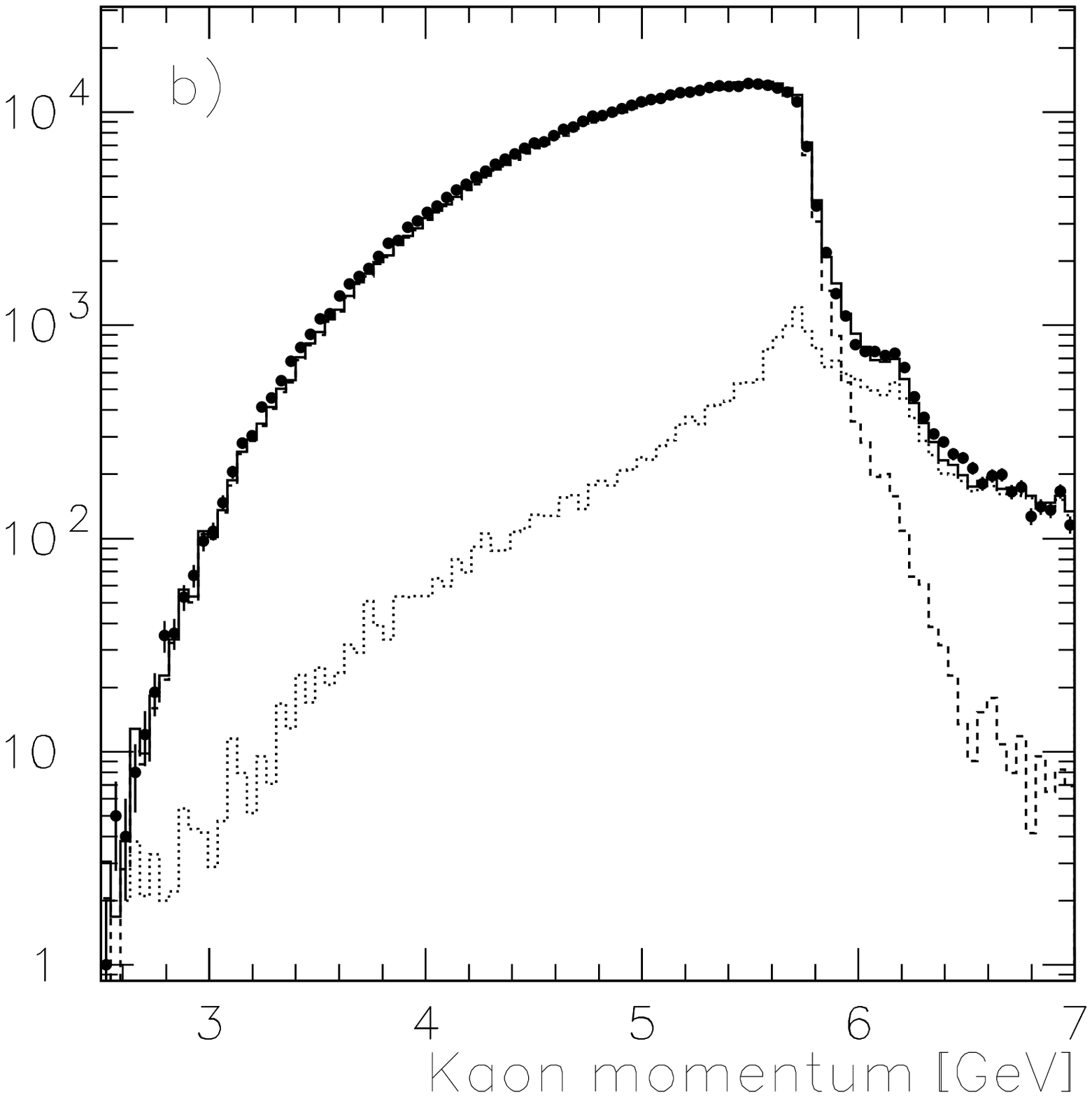}\centering
\includegraphics[width=0.33\linewidth]{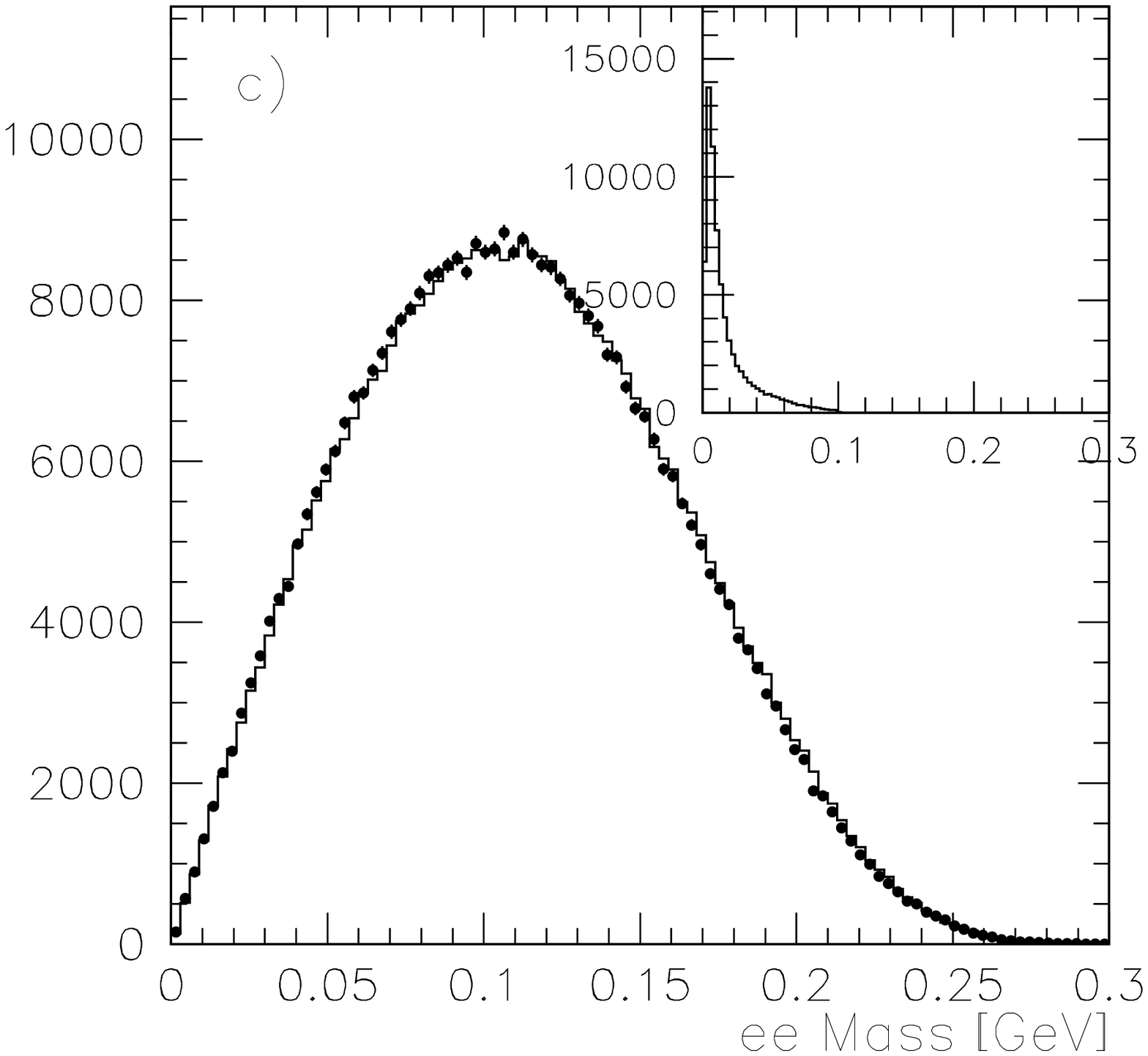}\centering
\caption[Background contributions:]
{Background contributions: the markers show the data while 
the solid histogram displays the Monte Carlo simulation. 
a) Three-pion invariant 
mass distribution for $K_{e4}$ candidate events, assigning a pion mass
to the positron. The small peak at the $K^+$ mass 
arises from $K_{\tau}$ events. b) Total momentum reconstructed from
the three charged track momenta. The solid histogram
is the sum of the Monte Carlo simulation of $K_{e4}$ events 
(dashed histogram) and 
the background from 2-1 accidentals (lower dotted histogram).
c) Electron-positron invariant mass $M_{ee}$ assigning electron mass 
to the reconstructed $\pi^-$ for $K_{e4}$ events. $K_{dal}$
events (insert) are characterized by low values of $M_{ee}$.} 
\label{fig:background}
\end{figure*}
Here we will mention only its main features. The detector was located
in a 6~GeV unseparated beam of approximately $1.5\times 10^7$ $K^+$
accompanied by about $3\times 10^8$ $\pi^+$ and protons per machine
spill of 1.6-2.8~s duration. About 6\% of the kaons accepted by the
beam line decayed in the 5~m long
evacuated decay volume. The decay products were separated by charge
and swept away from the beam by a first dipole magnet. Negatively 
charged particles were deflected to the left. A second dipole magnet 
sandwiched between four
proportional wire chambers (P1-P4) served 
as the spectrometer. The wire chambers, each consisting of four wire planes, 
were deadened in the region where the beam passed. This arrangement yielded 
a momentum resolution of $\sigma_P\simeq 0.003\,P^2$ GeV$/c$, where $P$, the 
momentum of the decay products in GeV$/c$, had a typical range of 0.6 to 3.5.
Pions and muons were distinguished from positrons and electrons 
using two \v{C}erenkov 
counters, C1 and C2, situated inside and behind the second dipole magnet,
and rendered insensitive in the beam region. 
Both \v{C}erenkov 
counters, when filled with CH$_4$ at atmospheric pressure, yielded 
on average seven photoelectrons, and hence insured an electron 
identification probability greater than 99 \%. An electromagnetic calorimeter
of the Shashlyk design~\cite{atoyan92}, located downstream of P4 further 
aided the separation of the positrons from other charged decay products. 
It consisted of 30 modules 
in the horizontal and 20 modules in the vertical direction, but for the beam
region, where $6\times 3$ modules were absent. Module size was 
11.4~cm high
and 11.4~cm wide perpendicular to the beam direction and 15 radiation
length deep. The calorimeter was followed by an array of 12 muon chambers, 
separated by iron planes, employed to discriminate pions against muons.
Four hodoscopes were added to the detector for trigger purposes. The
A-hodoscope was situated just upstream of the calorimeter, the B- and
C-hodoscopes were embedded in the muon stack, and the D-hodoscope was 
located between the first two proportional wire chambers. The detector 
was completed by a pixel counter, installed just upstream of the
decay volume, which measured the position of
the incoming kaons. This device consisted of an array of 12 (horizontally)
by 32 (vertically) scintillating pixels, each with an area of 
$7\times 7$~mm$^2$.

Table~\ref{tab:resolution} summarizes the resolution of the apparatus
in the five variables required to describe the kinematics of the
$K_{e4}$ decay.

\begin{table}[htb]
\begin{center}
 \begin{ruledtabular}
\begin{tabular}{lll}
Variable       & FWHM  \\ 
\hline
$s_{\pi}$      & 0.00133 &GeV$^2$    \\ 
$s_{e}$        & 0.00361 & GeV$^2$    \\ 
$\theta_{\pi}$ & 0.147 & rad    \\ 
$\theta_{e}$   & 0.111 & rad   \\ 
$\phi$         & 0.404 & rad   \\ 
\end{tabular}  
\end{ruledtabular}
\end{center} 
\caption[]{Experimental resolutions for the five kinematic variables
used in the analysis.
\label{tab:resolution}}
\end{table} 
\subsection{Trigger requirements}
\label{sec:trigger}

\begin{figure*}[htb]
\includegraphics[width=0.33\linewidth]{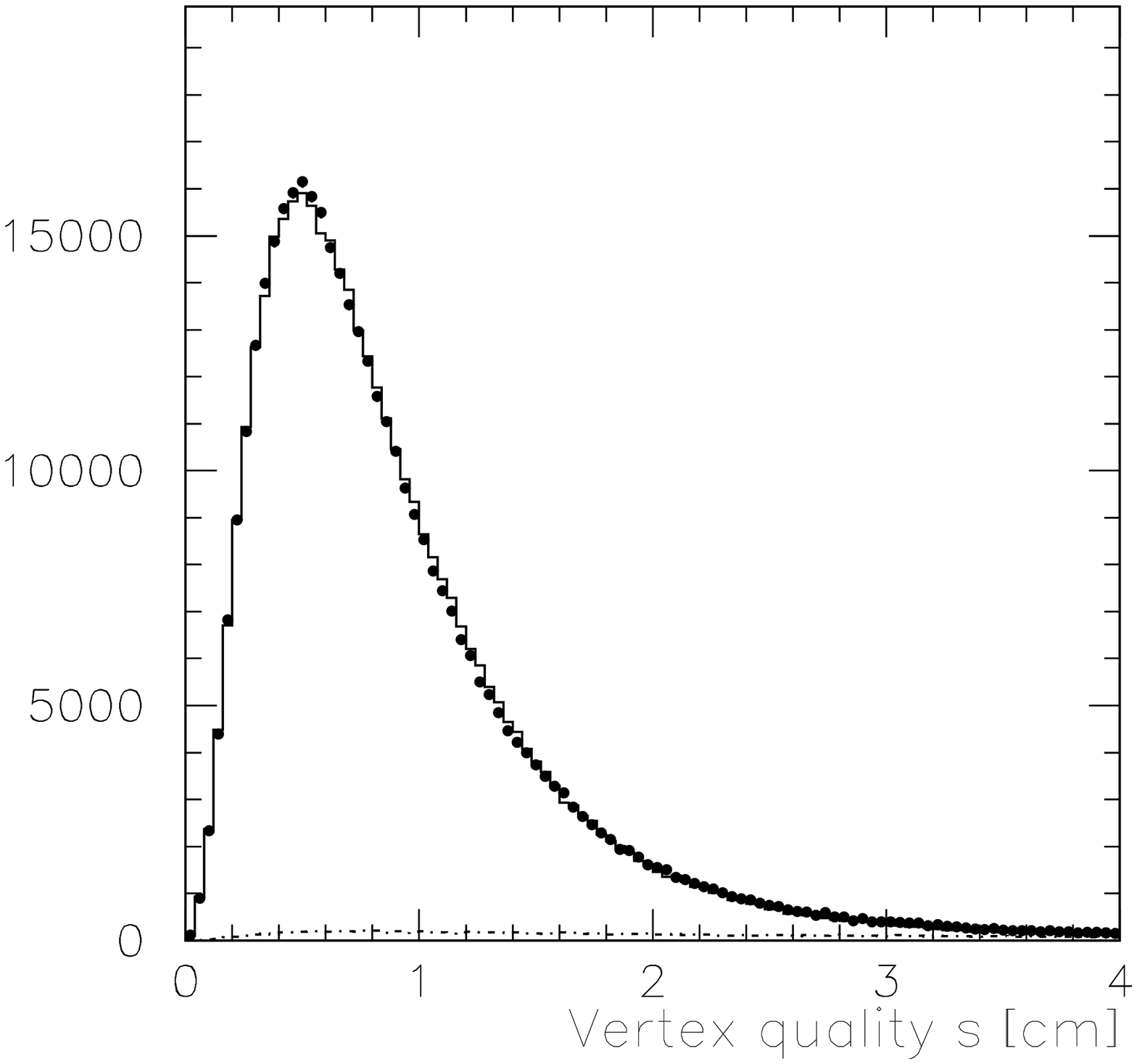}\centering
\includegraphics[width=0.33\linewidth]{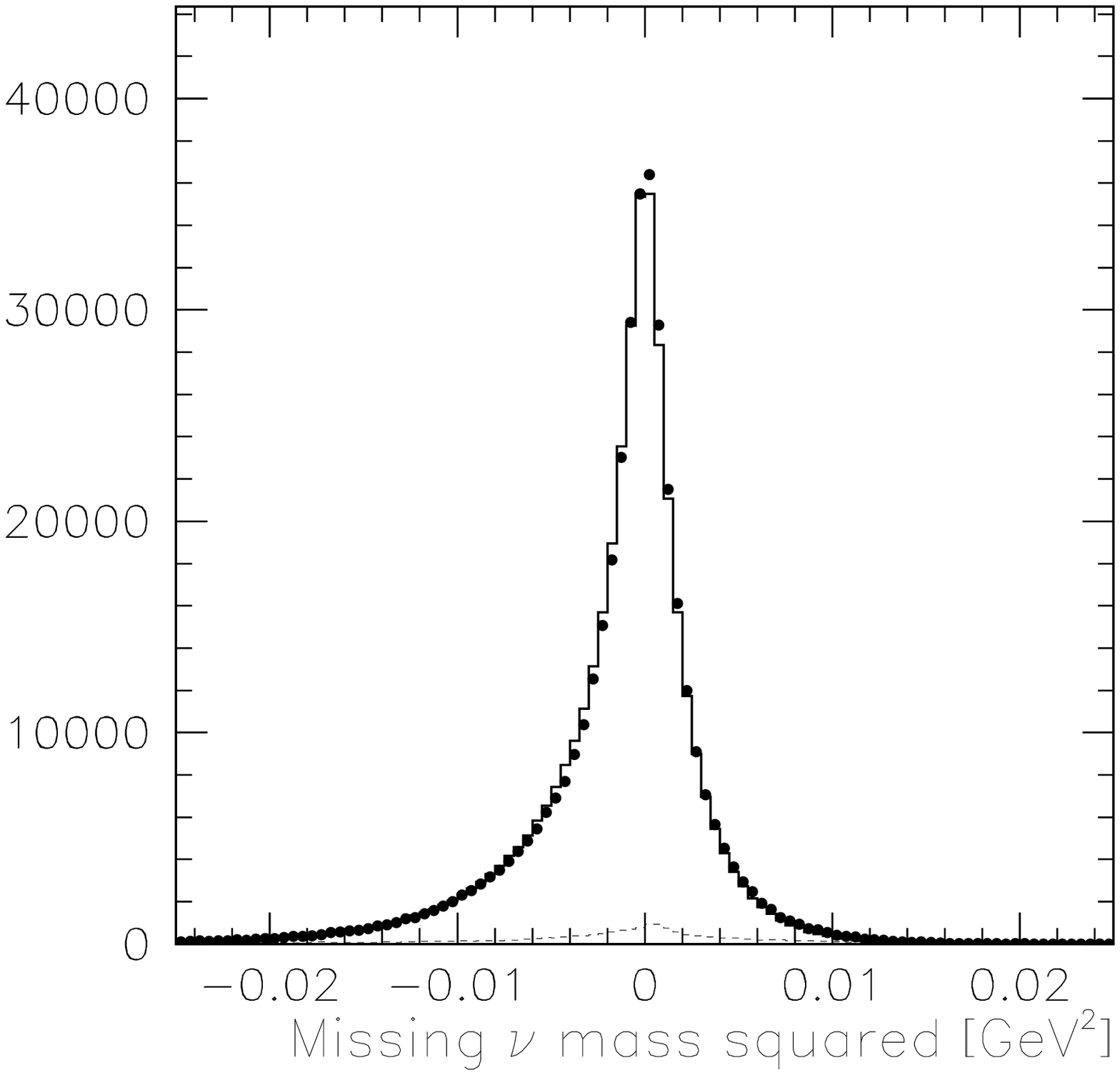}\centering
\includegraphics[width=0.33\linewidth]{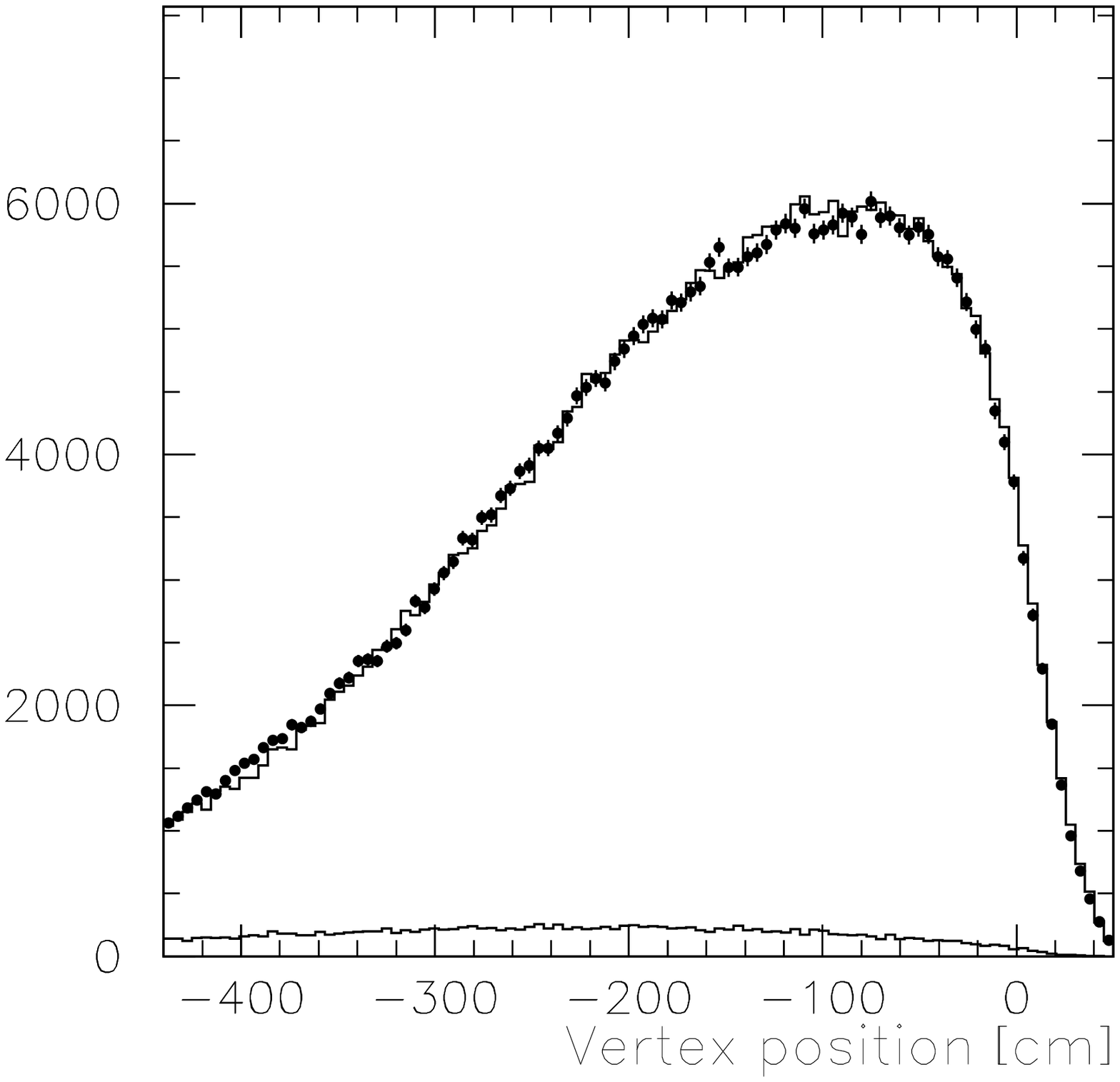}\centering
\caption[Monte Carlo control plots]
{Comparison of the Monte Carlo simulation (histogram) with data
(markers with error bars). Left: distance of closest approach $s$ to the
common vertex for the three charged tracks; center: missing neutrino mass
squared; right: distribution of decay vertices along the beam direction
$z$ ($z=0$ at the entrance of the first dipole magnet).
The dashed histograms show the background contributions.
\label{fig:mccomp}}
\end{figure*}

The trigger was designed as a multilevel structure
with increasing sophistication. The lowest trigger level (T0) indicated 
the presence of three charged particle tracks, two on the right 
and one on the left side, each signaled by a coincidence 
between the A-counter and the corresponding calorimeter module directly 
behind it (A$\cdot$SH). For each combination of coincidences on 
the right only a limited, kinematically acceptable region on the left 
was allowed. To insure that the trigger resulted from particles coming from 
the decay volume, at least one coincidence on both sides between the 
D-counter and A$\cdot$SH was required. The next trigger level 
(T1) demanded the presence of a positron in order to reject events from
the $K^+\rightarrow \pi^+\pi^+\pi^-$ ($K_{\tau}$) decay, and  
dismissed all events with evidence for the presence of an electron
to eliminate events from $K^+\rightarrow\pi^+\pi^0$ 
($\pi^0\rightarrow e^+e^-\gamma$, $K_{\rm dal}$) decay, both rather
common decay modes. Consequently,
this trigger level required signals in both \v{C}erenkov
counters on the right (corresponding to at least 2.5 photoelectrons) 
and vetoed 
all events with a signal in either \v{C}erenkov counters on the left
(at least 0.25 photoelectrons).
The final trigger level (T2) rejected events with a high occupancy in the wire 
chambers, most likely caused by noise in the read-out electronics. It did 
not reject many events, but the ones it rejected would have required an 
exceedingly 
large amount of computer time in the reconstruction. 
In addition to $K_{e4}$ candidates, a few prescaled monitor
triggers were also recorded, e.g. a minimum bias trigger
(T0 without the T1 requirement) dominated by accidentals and $K_{\tau}$ events,
and a trigger sensitive to $K_{dal}$ events, used to
check the \v{C}erenkov counter efficiency~\cite{appel02}.

\section{$K_{e4}$ event selection and analysis}
\label{sec:eventsample}

\subsection{Reconstruction}
\label{sec:eventreconstruction}
The kinematic reconstruction of an event, described in detail
in~\cite{appel02}, proceeded as follows: In the first step raw wire 
hits in the proportional chambers were combined to space points,
requiring signals in at least three of the four wire planes in a chamber.
Then the space points were combined to tracks. A track was found if at 
least three chambers contributed with a space point each. Next, employing 
a measured map of the magnetic field in the dipole magnets, the momenta 
of the tracks were fitted. For events with at least three reconstructed 
tracks, a fitting algorithm, again utilizing the field map, determined the 
decay
vertex as the position from which the distance $s$ to the three tracks 
was minimal. 
For events containing more than three tracks, the combination that 
produced the lowest $s$ was tagged as the most probable set of track 
candidates 
from kaon decay. Finally, the kaon direction was 
obtained from the hit in the pixel 
counter and the vertex. The kaon momentum could then be fitted by tracing 
the kaon back through the beam line to the production target 27.5~m upstream 
 of the decay tank. In the last reconstruction step the particle 
identification 
information was assigned to the tracks found. 

\subsection{Selection}
\label{sec:eventselection}
$K_{e4}$ candidates had to pass the following selection criteria:
a vertex within the decay
tank of acceptable quality $s$, a momentum  
reconstructed from the three daughter particles below 
the beam momentum, a timing spread 
between the signals caused by the tracks in the A-hodoscope and the 
calorimeter consistent with the resolution of 0.5~ns.
Finally we required an unambiguous identification of the $e^+$, 
assured by light in the appropriate photomultiplier tubes
in both \v{C}erenkov counters and an energy loss in the calorimeter 
consistent with the momentum of the track, and
 of the $\pi^-$, secured by the absence of a signal above the noise 
in the \v{C}erenkov counters and an energy loss in the calorimeter 
consistent with that of a minimum ionizing particle or a hadron shower . 
The cuts described above ensured $K_{e4}$ events of good quality, but the
resulting event sample still contained a considerable amount of
background events.

\subsection{Backgrounds}
\label{sec:backgrounds}
The major background contributions came from  $K_{\tau}$ decay 
and accidentals. A $K_{\tau}$ could fake a $K_{e4}$ by either (1) a
misidentification of one of the $\pi^+$ as a positron
due to $\delta$-rays, noise
in the photomultiplier tubes or the presence of an additional 
parasitic positron, or (2) a decay of a $\pi^+$ directly or via a 
$\mu^+$ into an $e^+$. The dominating accidental background arose from
combinations of a $\pi^+$ and a $\pi^-$ originating from a 
$K_{\tau}$ decay with a 
positron from either the beam or from a $K_{dal}$ decay
(2-1 accidental from $K_{\tau}$).

To reject background from $K_{\tau}$ decay, we required that the
kaon reconstructed from the three charged daughter particles did
not track back to the target, using the fact that the reconstruction
for $K_{e4}$ is incomplete due to the undetected neutrino. The remaining
$K_{\tau}$ background can be made 
visible by plotting the $K_{e4}$ candidates under the $K_{\tau}$
hypothesis, i.e. assigning to the positron a pion mass. The $K_{\tau}$
background appears as a narrow peak sitting on the broad distribution
originating from $K_{e4}$ decays, as seen in 
Fig.~\ref{fig:background}a.

Accidentals of the 2-1 type from $K_{\tau}$ are characterized by: 
(1) the positron 
track tends to be out of time in the A-hodoscope and the calorimeter 
compared with the two pion tracks; (2) the distance of closest approach 
between the positron track and each pion track is typically larger than 
the distance between the two pions; (3) the position of the vertex along
the beam axis tends to be more upstream in $K_{\tau}$ and hence also in
2-1 accidentals from $K_{\tau}$ compared 
with $K_{e4}$, due to smaller average transverse momentum; 
(4) in the calorimeter more clusters of 
energy are found, due to the possibility of two decays in the 
same time window. These characteristics were used to construct
a likelihood function in order to suppress 2-1 accidentals. 
The remaining background
can be exposed by inspecting the distribution of the total visible momentum
in the event, reconstructed from the sum of the three charged particle momenta.
Accidentals of the 2-1 type
display a large tail above the beam momentum, as is demonstrated
in Fig.~\ref{fig:background}b. The agreement between data and the sum of
Monte Carlo and background indicates that this background is well
understood. For the background simulation we used
$K_\tau$ monitor events with a fourth accidental positron track.
The uncertainty in the evaluation of this background
under the signal region below the beam momentum yields
the largest contribution to the systematic error of the
background estimate.

The excellent particle identification capabilities of our apparatus
reduce the background originating from $K_{dal}$ decay, where the
$e^-$ gets misidentified as a $\pi^-$, to a negligible level.
This can be made evident by plotting the invariant
mass $M_{ee}$ of the electron-positron pair,
assigning the electron mass to the $\pi^-$ (Fig.~\ref{fig:background}c).
This distribution shows no enhancement at the low
values of $M_{ee}$ characteristic for $K_{dal}$ events.

Table~\ref{tab:background} summarizes the background rates.
\begin{table}[htb]
\begin{center}
\begin{ruledtabular} 
\begin{tabular}{ll}
Background          & Fraction  \\ 
\hline
$K_{\tau}$ with $\pi^+$ misidentification & $(1.3\pm 0.3)\cdot 10^{-2}$    \\ 
$K_{\tau}$ with $\pi^+\rightarrow e^+\nu_e$ & $(3.5\pm 0.2)\cdot 10^{-3}$ \\ 
$K_{\tau}$ with $\pi^+\rightarrow \mu^+\nu_\mu$ and 
$\mu^+\rightarrow e^+ \nu_e \overline{\nu}_\mu$& $(2.6\pm 0.3)\cdot 10^{-3}$\\
$K^+\rightarrow \pi^0 \pi^+$$^{[a]}$ & $(2.5\pm 0.6)\cdot 10^{-5}$\\
$K^+\rightarrow \pi^0 e^+\nu_e$$^{[a]}$ & $(0.4\pm 0.1)\cdot 10^{-5}$\\
$K^+\rightarrow \pi^0 \mu^+\nu_\mu$$^{[a]}$  & $(0.4\pm 0.1)\cdot 10^{-5}$\\
$K^+\rightarrow \pi^+ \pi^0\pi^0$$^{[a]}$ & $(0.3\pm 0.1)\cdot 10^{-5}$\\
1-1-1 accidentals & $(0.9\pm 0.4)\cdot 10^{-4}$ \\
2-1 accidentals from $K_{\tau}$ & $(2.4\pm 1.2)\cdot 10^{-2}$ \\
2-1 accidentals from $K_{dal}$ & $(0.9\pm 0.4)\cdot 10^{-3}$ \\
\end{tabular}  
\end{ruledtabular} 
\end{center} 
\caption[Background contributions]{Compilation of fraction of 
background events.
1-1-1 accidentals: accidental
combinations of two independent pion tracks and a positron track;
2-1 accidentals: combinations of two pions from
a $K_{\tau}$ with an accidental positron or
combinations of a $\pi^+$ and a positron from a $K_{dal}$ decay
with an accidental $\pi^-$;
$^{[a]}$ $\pi^0\rightarrow e^+e^-\gamma$ and $e^-$ misidentification. 
\label{tab:background}}
\end{table} 

\subsection{Final sample}
\label{sec:finalsample}
After applying the event selection criteria described above,
406,103 events remained, of which we estimate $388,270\pm 5,025$
to be  $K_{e4}$ events. This corresponds to an increase in
statistics by more than a factor of 10 compared with previous 
experiments.

\section{Monte Carlo simulation}
\label{sec:montecarlo}
A good Monte Carlo simulation of the detector is a necessary ingredient
for the analysis of the decay distributions and the determination of the
absolute branching ratio. This simulation starts with the kaon beam
at the upstream end of the decay tank with a spatial and momentum 
distribution deduced from our ample supply of $K^+_\tau$ monitor
events, for which the incident $K^+$ can be fully reconstructed. 
The $K^+$ is then allowed to decay in a preselected mode along 
its trajectory in the decay tank.
To model the physics of the $K_{e4}$ decay, initial values of the matrix 
elements were chosen in accordance with the ChPT analysis at the one loop
level ~\cite{bijnens90,riggenbach91} of the Geneva-Saclay 
experiment~\cite{rosselet77}. Radiative corrections 
are included following Diamant-Berger~\cite{diamant76} (see also
Sec.~\ref{sec:fitform factors} below). 
For the decay modes $K_\tau$ and $K_{dal}$, needed for the determination 
of the branching ratio and the evaluation of the background, we use the
matrix elements given in Ref.~\cite{ktaumat} and \cite{kdalmat}.
The detector response is  
handled with a GEANT-based~\cite{geant} simulation of the E865 
apparatus, and the simulated events are processed through the same 
reconstruction and selection programs as data events.
With these tools, we generated $81.6\cdot 10^6$ $K_{e4}$ events,
resulting in $2.9\cdot 10^6$ accepted 
events, about 7.5 times more than data events.
The quality of the simulation is
demonstrated in Fig.~\ref{fig:mccomp}, which displays the vertex quality 
$s$, the missing 
neutrino mass squared, and the position of the vertex along the
beam axis as examples. The vertex quality is a crucial
quantity in the
event reconstruction; the missing neutrino mass squared is 
sensitive to the resolution; and the vertex position depends on the
decay matrix element and detector acceptance.
The good agreement between data and
Monte Carlo indicates that ChPT describes the data well and that
our event selection procedure did not introduce a significant bias.
We also compare Monte Carlo with data distributions for the 
kinematically very distinct $K_{\tau}$ and $K_{dal}$ decays, getting
again a nice agreement (see, e.g.~\cite{appel02}). Furthermore, we 
find that the $K_{dal}$ branching ratio is consistent with the 
published value~\cite{groom00}, using $K_{\tau}$ as normalization channel.  
This underlines the good understanding of the geometrical acceptance and the 
efficiency of the various detector elements.

\section{Branching ratio}
\label{sec:branchingratio}
The $K_{e4}$ branching ratio was normalized with respect to the $K_{\tau}$ 
decay. As mentioned in Sec.~\ref{sec:trigger}, we collected $K_{\tau}$
events in a minimum bias trigger concurrently with $K_{e4}$ events. 
$K_{\tau}$ is the most common kaon decay with three charged particles in 
the final state, which strongly simplifies the selection of a clean sample of
events. To identify $K_{\tau}$ events, we require the
reconstruction of a vertex, as for $K_{e4}$, and the 
reconstruction of the kaon mass. With 
$BR(\tau)=5.59\pm0.05$ \%~\cite{groom00}, 
the $K_{e4}$ branching ratio $BR$ and the decay rate $\lambda$ are  calculated
as
\begin{eqnarray}
BR(K_{e4})&=&BR(K_\tau)\frac{N(K_{e4})A(K_\tau)}{N(K_\tau)A(K_{e4})}C\nn
N(K_{e4})\;[N(K_\tau)]&=& {\rm number\; of\;}K_{e4}\;[K_\tau]\;{\rm events}\nn
&=& 388,270\pm 5,025\;[1.487\cdot 10^9]\nn
A(K_{e4})\;[A(K_\tau)]&=& {\rm acceptance\; for\;}K_{e4}\;[K_\tau]\;
{\rm events}\nn
&=& 3.77\,\%\;[10.29\,\%]\nn
C&=& {\rm accidental\; veto\; correction } \nn
& = & 1.0312\pm 0.0022\ ,\label{eq:br1}
\end{eqnarray}
leading to 
\begin{eqnarray}
BR(K_{e4})&= & (4.109\pm0.008\pm0.110)\cdot 10^{-5} \nn 
\lambda (K_{e4})& = & (3321\pm 6\pm 89)\;{\rm s}^{-1}
\label{eq:br2}
\end{eqnarray}
The first error is statistical, the second is systematic. The result
is in good agreement with previous
experiments, as is evident from Table~\ref{tab:brold}.

\begin{table}[htbp]
  \begin{center}
    \begin{ruledtabular}
    \begin{tabular}{lrr}
Reference & \# of events & Branching ratio\\\hline
PDG~\cite{groom00}     &    & $(3.91\pm0.17)\cdot 10^{-5}$ \\
Rosselet {\it et al.}~\cite{rosselet77}& 30318  
& $(4.03\pm0.17)\cdot 10^{-5}$\\
Beier {\it et al.}~\cite{beier72} & 8141 & \\
Bourquin {\it et al.}~\cite{bourquin71}& 1609 & $(4.11\pm0.38)\cdot 10^{-5}$\\
Schweinberger {\it et al.}~\cite{schweinberger71}& 115
& $(3.91 \pm0.50) \cdot 10^{-5}$\\
Ely {\it et al.}~\cite{ely69} &  269   & $(3.26\pm0.35)\cdot 10^{-5}$\\
Birge {\it et al.}~\cite{birge65} & 69 & $(3.74\pm0.84)\cdot 10^{-5}$
    \end{tabular}
    \end{ruledtabular}
    \caption{$K_{e4}$ branching ratios measured in older experiments.}
    \label{tab:brold}
  \end{center}
\end{table}
The systematic errors are summarized in
Table~\ref{tab:brsys}. The dominant contributions are from the background
subtraction and \v{C}erenkov counter efficiencies. The error in the background
subtraction results from the uncertainty in the background rate for
2-1 accidentals from $K_{\tau}$, as mentioned in Sec.~\ref{sec:backgrounds}.
The efficiency of the \v{C}erenkov counters was determined using  $K_{dal}$
decays, collected with the special purpose \v{C}erenkov counter trigger
described in Sec.~\ref{sec:trigger}. The uncertainty results from 
the fact that $K_{dal}$ events populate phase space areas different from 
$K_{e4}$. This is mainly significant on the beam right side, where  2.5 
photoelectrons are required to identify a positron.
The branching ratio includes radiative $K_{e4}$ events, i.e. 
$K^+\rightarrow\pi^+\pi^-e^+\nu_e\gamma$, since no cut
on the missing neutral mass squared is made. 
Diamant-Berger~\cite{diamant76} found that the ratio of radiative to 
non-radiative $K_{e4}$ events for photon energies above 30 MeV is
only $(1.0 \pm 0.5) \%$. A small fraction of these, which lead to
an additional cluster in the calorimeter could be rejected, because 
the number of clusters is used in the likelihood function for background
rejection.

\begin{table}[htb]
\begin{center}
\begin{ruledtabular} 
\begin{tabular}{ll}
Sources:                   & $\sigma_{BR}/BR$ \\[1mm]
\hline
Background subtraction     & 0.012 \\
$K_{\tau}$ prescale factor & 0.0076 \\
Magnetic field map         & 0.005 \\
\v{C}erenkov counter inefficiencies  & 0.015 \\
PWC efficiencies           & 0.006 \\
Fiducial volume            & 0.005 \\
Track quality              & 0.0022 \\
Vertex reconstruction      & 0.0016 \\
Z-position of vertex       & 0.0012 \\
Tracking back to target    & 0.0019 \\
Timing cuts                & 0.0020 \\
$e^+$ identification in the calorimeter & 0.0007 \\
$\pi^-$ identification    & 0.0011 \\
2-1-accidental likelihood   & 0.0006 \\
$K_{e4}$ matrix element (statistics)    & 0.006 \\
$K_{\tau}$ mass resolution  & 0.0081 \\
$K_{\tau}$ branching ratio & 0.009  \\
\hline
Total (added quadratically) & 0.0268 \\
\end{tabular}
\end{ruledtabular} 
\end{center} 
\caption[Systematic errors]
{Systematic errors in the branching ratio measurement.
\label{tab:brsys}}
\end{table}

\section{Fits to the decay distributions}
\label{sec:fitform factors}

In pursuing the goal of determining the form factors and
$\pi\pi$ scattering phase shifts, three different approaches
have been followed, which have been outlined in Sec.~\ref{sec:theory_parametrisation}. The $K_{e4}$ form factors $F$, $G$, and $H$, and
the phase shift $\delta$ can be directly extracted for a conveniently
chosen grid of bins in the kinematic variables. This approach makes 
no assumption on the analytical behavior of these quantities.
In the second approach, the parametrisation of Eq.~\ref{equ:amoros} is
used and the phase shifts are related to the
two scattering lengths using Eq.~\ref{equ:schenk}. 
This allows use of the whole data sample
in a single fit. Finally, either Eq.~\ref{equ:universalcurve} or
Eq.~\ref{equ:universalcurvegilberto} can be used in addition, 
reducing the number of parameters by one. The statistical method which we
describe below is the same for all three approaches.
\begin{figure*}[hbt]
\begin{center}
\includegraphics[width=120mm]{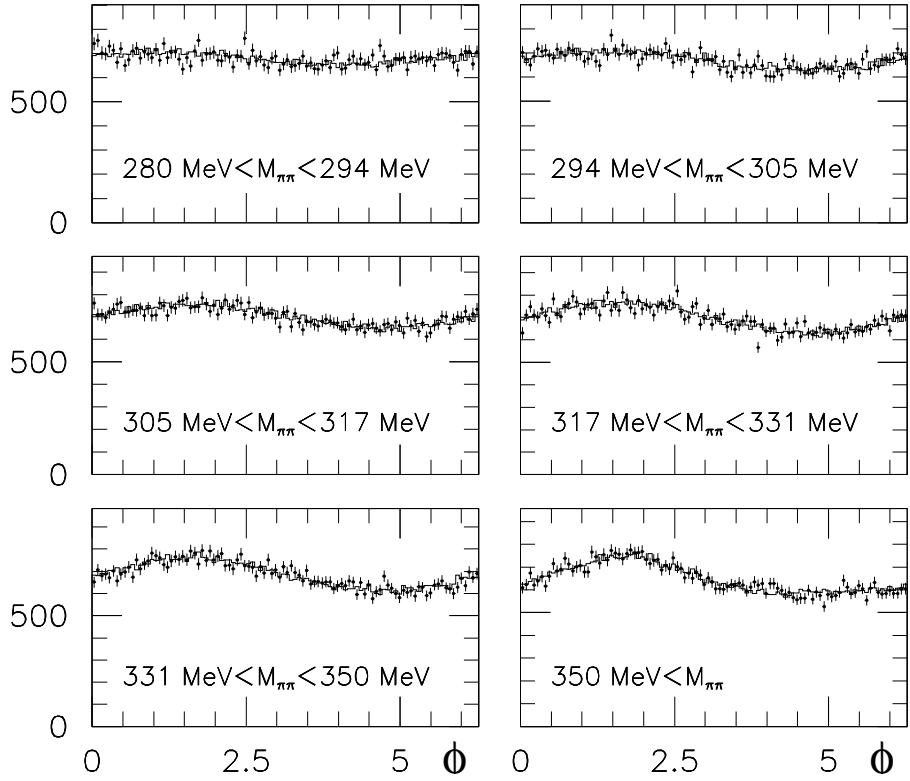}
\end{center}
\caption[Phifits]{$\phi$ distributions for the six bins in $M_{\pi\pi}$.
The markers with error bars represent the data, the histogram
the modified Monte Carlo distribution after the fit.\label{fig:phifits}}
\end{figure*}
\begin{table*}[htb]
\begin{center}
\begin{ruledtabular} 
\begin{tabular}{lccc}
$M_{\pi\pi}$, $<M_{\pi\pi}>$ (MeV)    & 280-294, 285.2 
& 294-305, 299.5 & 305-317, 311.2  \\ 
\hline
$F$  & $5832 \pm 13\pm 80$ (-26) & $5875 \pm 14\pm 83$ (+34) 
                      & $5963 \pm 14\pm 90$ (+44)\\
$G$  & $4703 \pm 89\pm 69$ (+22) & $4694 \pm 62\pm 67$ (+27) 
                      & $4772 \pm 54\pm 70$ (+34)\\
$H$  & $-3740 \pm 800 \pm 180 $ (-59) & $-3500 \pm 520 \pm 190$ (-50) 
                      & $-3550 \pm 440 \pm 200 $ (-167) \\
$\delta=\delta^0_0-\delta^1_1$
   &$-16 \pm 40 \pm 2$ (+0.5) & $68\pm 25\pm 1$ (-0.4) 
                      & $134\pm 19\pm 2$ (-1.3)\\
\hline
$\chi^2/\mbox{ndf}$   & 1.071           & 1.080           & 1.066   \\
\hline\hline
$M_{\pi\pi}$, $<M_{\pi\pi}>$ (MeV)    & 317-331, 324.0 & 
331-350, 340.4 & $>350$, 381.4\\ 
\hline
$F$  & $6022 \pm 16\pm 94$ (+46) 
     & $6145 \pm 17\pm 96$ (+45) & $6196 \pm 20 \pm 83$ (+34) \\
$G$  & $5000 \pm 51\pm 82$ (+38)  
     & $5003 \pm 49\pm 83$ (+31) & $5105 \pm 50\pm 74$ (+31)\\
$H$  & $-3630 \pm 410 \pm 230$ (-177) 
     & $-1700 \pm 410 \pm 240$ (-160) & $-2230 \pm 480 \pm 330$ (-173) \\
$\delta=\delta^0_0-\delta^1_1$
     & $160\pm 17 \pm 2$ (+0.1) 
     & $212 \pm 15\pm 3$ (+0.2) & $284 \pm 14 \pm 3$ (+0.6) \\
\hline
$\chi^2/\mbox{ndf}$   & 1.103           & 1.093           & 1.034         \\
\end{tabular}  
\end{ruledtabular} 
\end{center} 
\caption[]{Form factors and phase shifts for the six bins in dipion
invariant mass $M_{\pi\pi}$
(in units of $10^{-3}$). $<M_{\pi\pi}>$ refers to the centroid of the
bin.
The number of degrees of freedom for each fit is 4796. 
The first errors  are statistical, the
second  systematic. The fourth quantity, which is in parentheses, indicates the
shift of the central value of the parameter which resulted from the 
application of the radiative corrections. 
$F$, $G$ and $H$ given here are the moduli of the complex
form factor defined in Eq.~\ref{equ:amoros}.  
\label{tab:formfactorbin}}
\end{table*} 

\subsection{Data treatment}
\label{sec:theoretical_aspects}

The experimental distributions must be fit to Eq.~\ref{equ:decayrate}, 
taking into account the acceptance and resolution of the apparatus, 
with the form factors and phase shifts as free parameters.
Following the recommendations by Eadie~\cite{eadie71}
we select equi-probable bins for each kinematic variable,
namely six bins in $s_\pi$, five in $s_e$, ten in 
$\cos\theta_\pi$, six in $\cos\theta_e$, and 16 bins in $\phi$.
With a total of 28,800 bins there are on average 13 events in each
bin.

Following the procedure used by the Geneva-Saclay 
experiment~\cite{diamant76,rosselet77}, we minimize a $\chi^2$ function 
defined as:
\begin{eqnarray}
\chi^2 & = & 2\sum_j n_j 
           \ln\left[\frac{n_j}{r_j}\left(1-\frac{1}{m_j+1}\right)\right] \nn
       & + & 2\sum_j(n_j+m_j+1)\ln\left[\frac{\textstyle 1+
       \frac{\textstyle r_j}{\textstyle m_j}}
       {\textstyle 1+\frac{\textstyle n_j}{\textstyle m_j+1}}\right]\ .
\end{eqnarray}
where the sum runs over all bins. $n_j$, $r_j$ and $m_j$ are
the number of data events, expected events and generated Monte Carlo events 
in bin $j$, respectively. This $\chi^2$ is
deduced from the probability 
\begin{displaymath}
P(n,m,r)=\int_0^\infty\int_0^\infty e^{-u}\frac{u^n}{n!}
         e^{-v}\frac{v^m}{m!}\delta(u-\frac{r}{m}v)du\ dv
\end{displaymath}
and takes into account the limited number of Monte Carlo events. It
reduces to the more familiar expression
\begin{displaymath}
\chi^2=\sum_j\left(2(r_j-n_j)+2n_j\ln(n_j/r_j)\right)
\end{displaymath}
for large  $m_j$.

The expected number of events $r_j$ is calculated to be
\begin{equation}
r_j = Br(K_{e4})\frac{N^K}{N^{MC}}\sum \frac{J_5(F,G,H)^{new}}
           {J_5(F,G,H)^{MC}} \co
\label{equ:fitmethod}           
\end{equation}
where the sum runs over all Monte Carlo events in bin $j$. 
$N^K$ is the number of $K^+$ decays derived from the number
of $K_\tau$ events. $N^{MC}$ is the number of generated events. 
$J_5(F,G,H)^{MC}$ (Eq.~\ref{equ:i}) is evaluated at the
relevant set of kinematic variables for the simulated event  
with the form factors $F$, $G$, and $H$ calculated at $q=q^{MC}$.
$J_5(F,G,H)^{new}$ is evaluated with the same kinematic set and $F$, $G$, $H$
recalculated from the parameters of the fit. 
Thus, we apply the parameters on an event by event basis, and
at the same time, we divide out a possible bias caused by the matrix
element, making the fit independent of the ChPT ansatz  
used to generate the Monte Carlo events. 

\subsection{Fit of the decay rate in multiple bins in $s_\pi$}
\label{sec:multiple_fits}

For the fit in multiple bins two further assumptions are being made,
namely that the form factors do not depend on
$s_e$ and that the form factor $F$ contributes to $s$-waves only.
This is equivalent to setting
$f_e$, $g_e$ and $\tilde{f}_p$ equal to zero in the parametrisation of
Ref.~\cite{amoros99}. The validity of these assumptions will be discussed in
Sec.~\ref{sec:form_factors_with_universal_curve} below. Hence 
the four parameters $F$, $G$, and $H$ and $\delta\equiv\delta_0^0-\delta_1^1$
are fit for each of the six bins in $M_{\pi\pi}=\sqrt{s_\pi}$.
Table~\ref{tab:formfactorbin} summarizes the results. Figure~\ref{fig:phifits}
shows the $\phi$ distribution for each of the bins, which illustrates the
high quality of the fit.

The centroids $<M_{\pi\pi}>$ of the
bins are estimated following the recommendations by Lafferty
and Wyatt~\cite{lafferty95}.
The dominant systematic error
for $F$, $G$, and $H$ has the same origin as that of
the branching ratio measurement. 
The major contributions to the systematic error of $\delta$ are
the subtraction of the background, and resolution effects, i.e.
deviations between the original and reconstructed kinematics.

\begin{figure}[htb]
\includegraphics[width=0.85\linewidth]{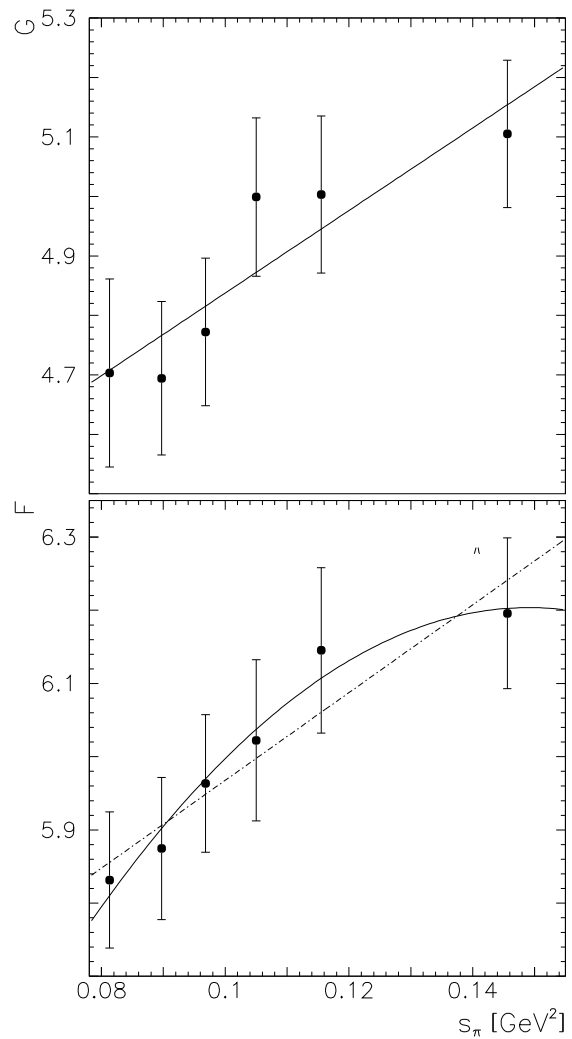}\centering
\caption[Phase shift difference]
{$s_\pi$ dependence of form factors $F$ and $G$.
\label{fig:formfactors}}
\end{figure}

We have also included the full magnitude of the radiative
corrections in the systematic error. As mentioned above in 
Sec.~\ref{sec:montecarlo}, we have calculated these 
corrections using formulae given
in Ref.~\cite{diamant76,diamant76a} based on the work of Neveu and
Scherk~\cite{neveu68}. Basically one has to consider two types of radiative
corrections, those where a real photon is radiated by one of the
charged particles involved in the decay and those where a virtual photon is
exchanged between two charged particles. The former are dominated by 
inner bremsstrahlung in particular of the positron~\cite{diamant76}, as
e.g. experimentally determined in the related decay 
$K^0_{Le3}\rightarrow\pi^\pm e^\mp\bar{\nu}_e(\nu_e)$~\cite{alavi01}.
The Low theorem~\cite{low58} insures that off-shell effects appear
only in second order and hence modifications of the hadronic form factors
are expected to be negligible. The Coulomb interaction of the charged
particles in the decay, however, has noticeable effects, in particular 
its most important contribution, the mutual attraction of the
pion pair, as already observed in the Geneva-Saclay 
experiment~\cite{diamant76,rosselet77}. The repulsion or attraction
between the positron, kaon and the two pions, which we also included, 
is unimportant. As an example we have reproduced the
$\pi\pi$ Coulomb attraction below~\cite{diamant76a}, 
which we have used to reweight each event: 
\begin{eqnarray}
d\Gamma_T&=&d\Gamma_0(1+\alpha C)\ ,\\[1ex]{\rm where}&&\nn
C&=&\pi\frac{1+v^2}{2v}+\frac{2}{\pi}\ln(\frac{2E_m}{m_\pi})
\left(\frac{1+v^2}{2v}\ln(\frac{1+v}{1-v})-1\right)\nn
&&+\frac{1}{\pi}(\frac{2+v^2}{2v})\ln(\frac{1+v}{1-v})+
\frac{8A}{\pi}(\frac{1+v^2}{2v})-\frac{1}{4\pi}\ ,\nn[1ex]
{\rm and}&&\nn
A&=&\int_0^{0.5\ln((1+v)/(1-v))}z\coth z\,dz\nn
&=&\mathcal{L}_2(v)-\mathcal{L}_2(-v)-\frac{1}{2}
\left(\mathcal{L}_2(\frac{2}{1+v})-\mathcal{L}_2(\frac{2}{1-v})
\right)\ ,\nn
&&\mathcal{L}_2(x)\equiv-\int_0^x\frac{1}{y}\ln|1-y|dy\ .\nonumber
\label{equ:radcor}
\end{eqnarray}
Here $v$ is the velocity of the pions in the dipion centre-of-mass
system (in units of $c$), $\alpha$ the fine-structure constant, and
$E_m$  a cut-off energy fixed at 30 MeV. In all tables where
results are given
(Tab.~\ref{tab:formfactorbin}, \ref{tab:ffwithuniversal},
\ref{tab:ffwouniversal} and \ref{tab:ffhigherorder})
we have listed the effect of applying the radiative corrections
separately. While the form factors $F$ and $G$ and the phase shifts
$\delta$ are nearly unaffected, the form factor $H$ changes 
between 1.5 and 9.4 \%.

The small deviation of $\chi^2/\mbox{ndf}$ from
the expected value of one 
may reflect the discreteness of the background.
The number of background events which we add to
the generated events is smaller than the
number of bins, and the
background is distributed over almost the whole phase space. 
By using tighter cuts, which reduce the
background contributions by a factor of two,
we have confirmed  that the results
for the form factors and phase shifts remain unchanged.

\begin{figure}[htb]
\includegraphics[width=0.85\linewidth]{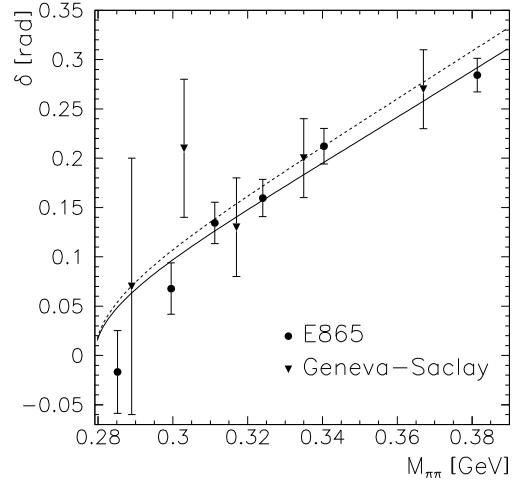}\centering
\caption[Phase shift difference]
{Phase shift difference $\delta$. The fits are given
by Eq.~\ref{equ:schenk} as a function of the scattering
length $a^0_0$. Solid line: this experiment; dashed line: 
Geneva-Saclay~\cite{rosselet77}}
\label{fig:scl}
\end{figure}

The results from Table~\ref{tab:formfactorbin} allow us to examine the
$s_\pi$ dependence of the form factors $F$, and $G$,
and of the phase  $\delta$, which are displayed in 
Figures~\ref{fig:formfactors} and \ref{fig:scl}. For the various fits
to these data, which we report below, the value of $\chi^2/$ndf
is always below one. Following Amor\'os and 
Bijnens~\cite{amoros99}, we fitted $F$ 
with a second degree polynomial, while a linear function suffices for $G$,
with the following results:
\begin{eqnarray}
f_s  =  5.77 \pm 0.10,& f_s^\prime= 0.95\pm 0.58,&
f_s^{\prime\prime}=-0.52\pm 0.61,  \nn
g_p  = 4.68\pm 0.09,& g_p^\prime= 0.54\pm0.20\ .&
\end{eqnarray}
Figure~\ref{fig:formfactors} also shows the results of a linear fit:
$F(q)=F(0) (1+\lambda_F q^2)$. We found 
\begin{equation}
F(0)  =  5.83\pm0.08 \co \;\;  \lambda_F= 0.079\pm0.015\ ,
\end{equation}
where the error of $\lambda_F$ was calculated using only the relative
errors of $F$ in the six bins.
These results are in agreement with those of the Geneva-Saclay 
experiment~\cite{rosselet77}, namely
\begin{equation}
F(0)  =  5.59\pm 0.14 \co \;\;  \lambda_F= 0.08 \pm 0.02\ .
\end{equation}
In the latter analysis it was assumed that 
$\lambda_F=\lambda_G\equiv g_p^\prime/g_p$ holds, which is confirmed
by our analysis, albeit within large error limits.

Good agreement with the previous measurements~\cite{rosselet77} and
considerably improved precision is shown Fig.~\ref{fig:scl},
where the phase shift difference $\delta$ is plotted versus $M_{\pi\pi}=\sqrt{s_\pi}$. 
A fit using Eq.~\ref{equ:schenk} with relation Eq.~\ref{equ:universalcurve},
taking the central curve of the universal band with 
the six data points for $\delta$ leads to 
the following value of the scattering length:
\begin{equation}
a_0^0=0.229\pm0.015\;\;(\chi^2/\mbox{ndf}=4.8/5).
\label{equ:a00bin}
\end{equation}
The use of Eq.~\ref{equ:universalcurve} then implies $a_0^2=-0.0363\pm0.0029$.

\subsection{Fits to the whole data set}
\label{sec:form_factors_with_universal_curve}
In this section we list the results of various fits to
the whole data sample. A more detailed discussion and comparison
will follow in Sec.~\ref{sec:conclusion}
\begin{figure*}[htb]
\includegraphics[width=0.33\linewidth]{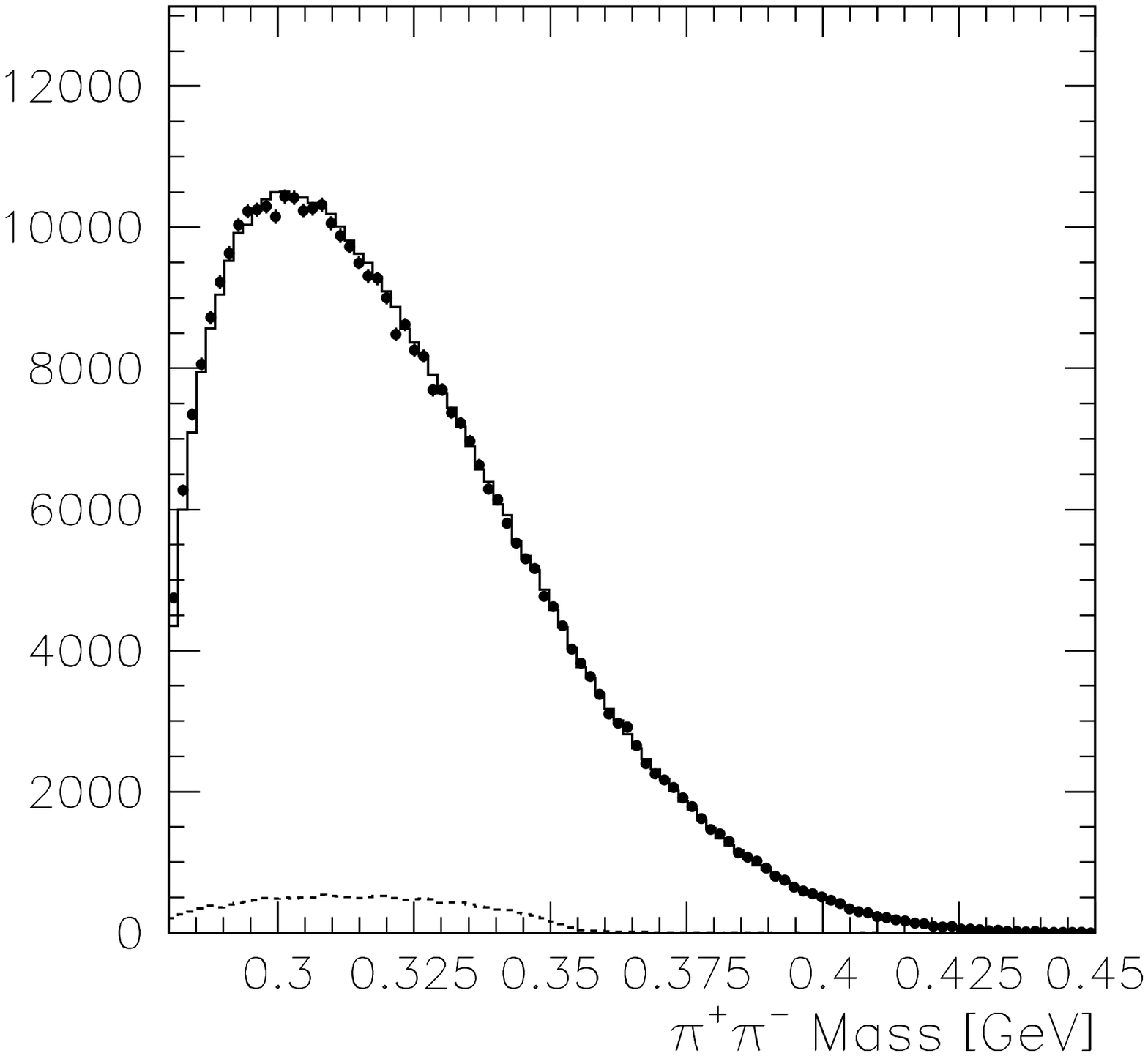}\hspace*{5mm}
\includegraphics[width=0.33\linewidth]{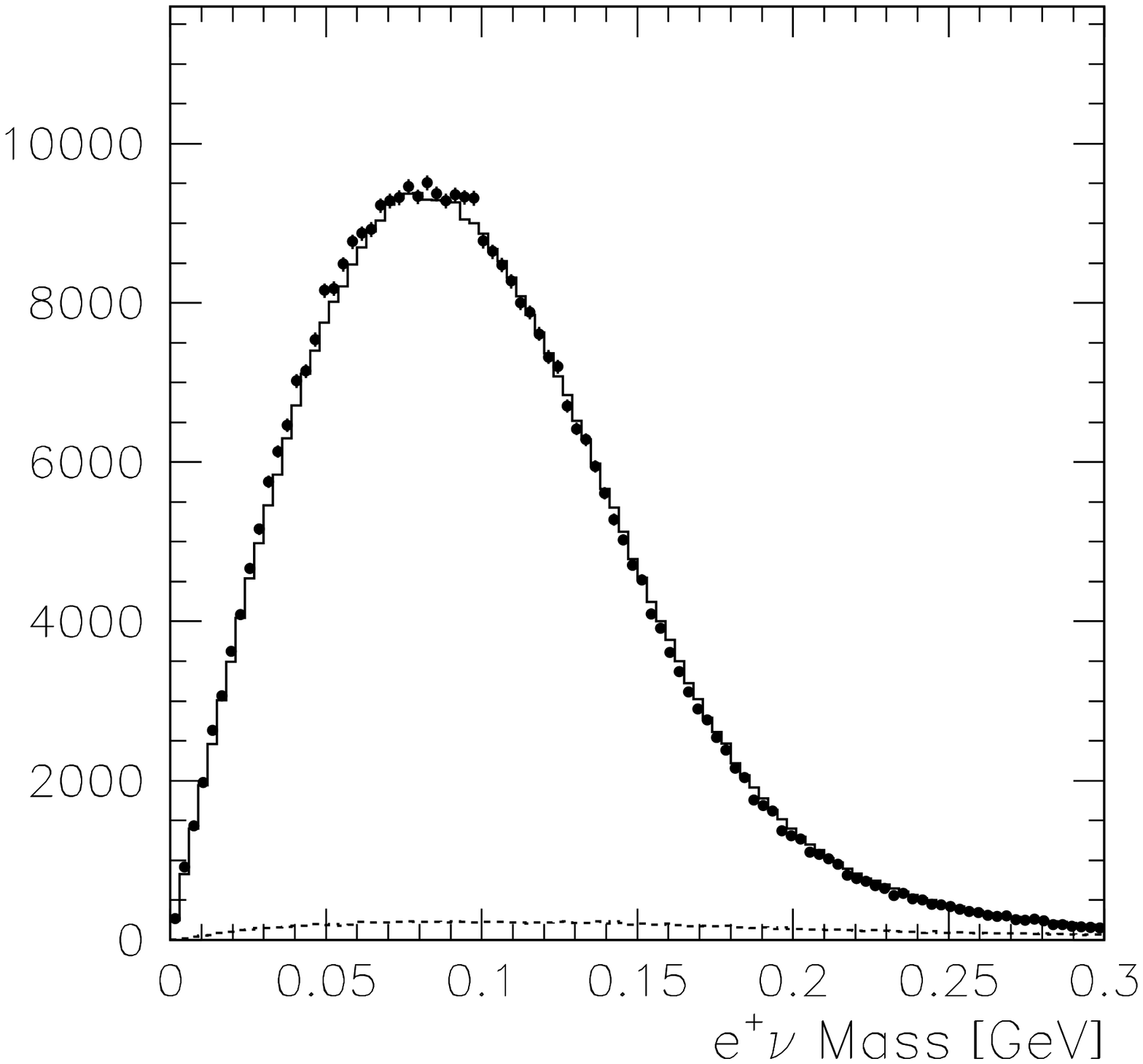}\centering
\\
\parbox{0.66\linewidth}{
\includegraphics[width=0.50\linewidth]{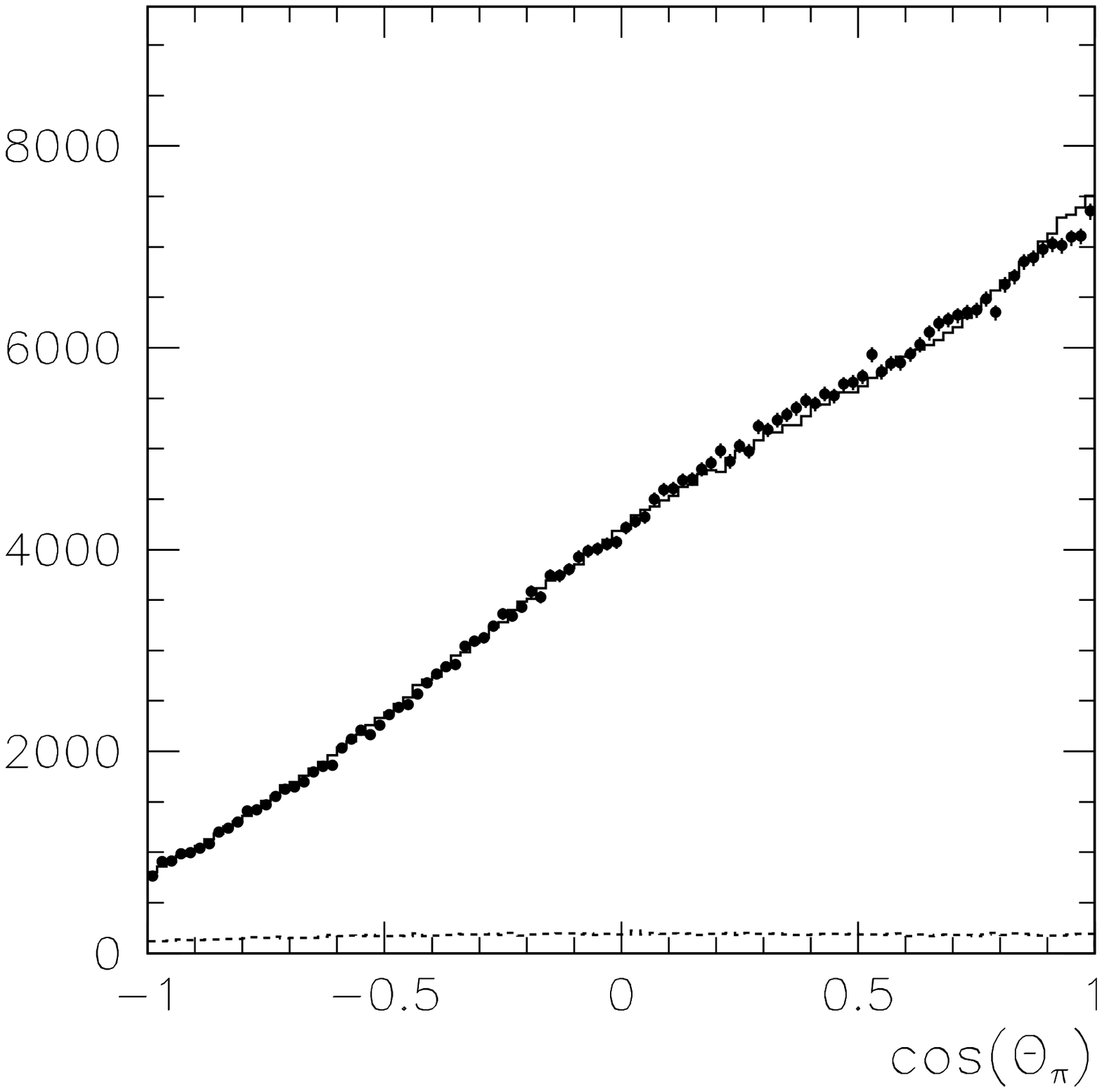}\hspace*{-2mm}
\includegraphics[width=0.50\linewidth]{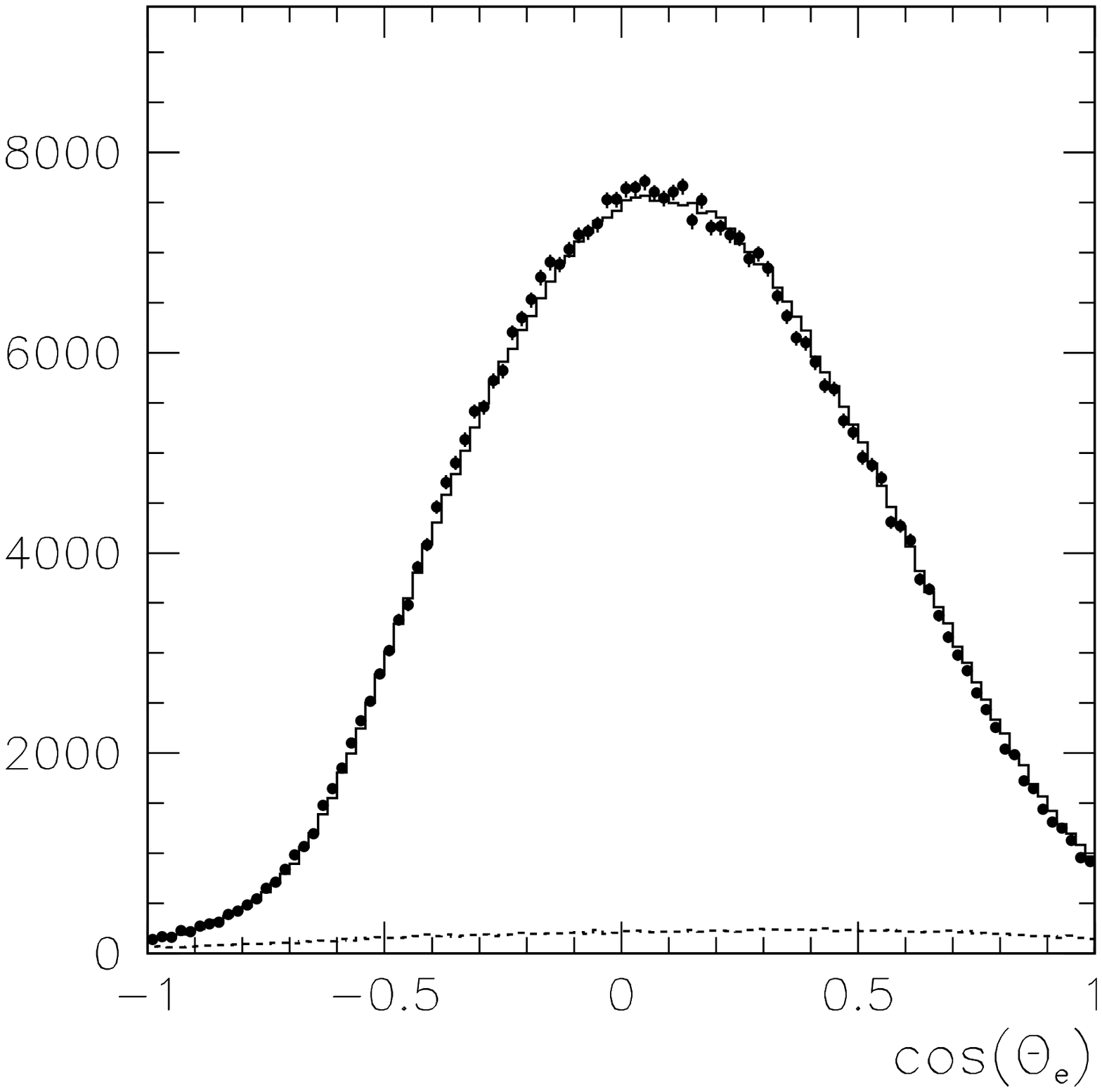}}\hspace*{-2mm}
\parbox{0.33\linewidth}{
\includegraphics[width=0.87\linewidth]{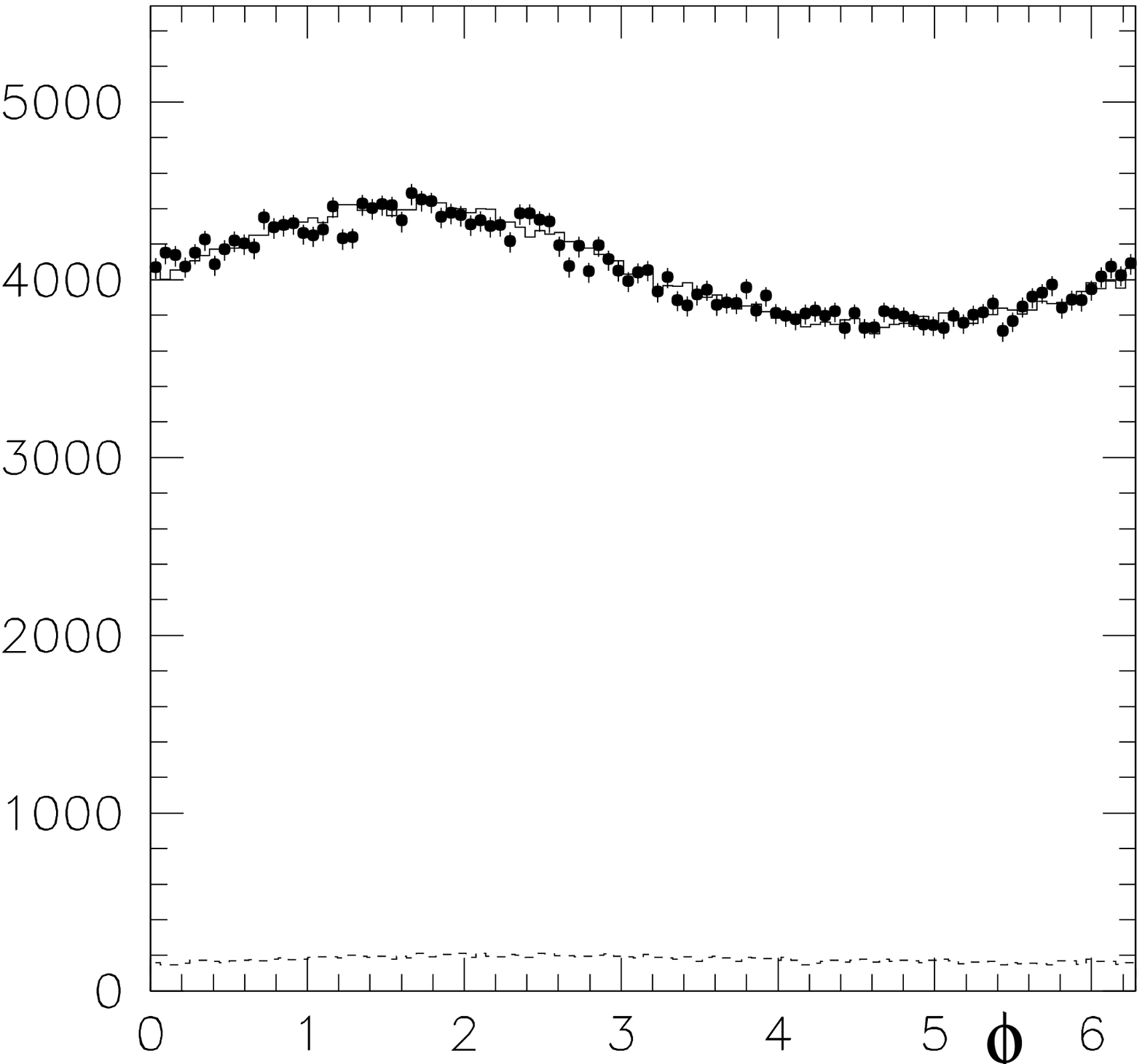}\vspace*{2mm}}\centering
\caption[Kinematics of the $K_{e4}$ decay.]
{Invariant masses and angles describing the $K_{e4}$ decay. 
The histograms are the Monte Carlo distributions while the
points with the error bars represent the data.
The dashed histograms show the background.\label{fig:mccompkin}}
\end{figure*}

If we substitute the phase shifts $\delta$ in Eq.~\ref{equ:amoros}
via Eq.~\ref{equ:schenk} and Eq.~\ref{equ:universalcurve}
or Eq.~\ref{equ:universalcurvegilberto} for the
relation between $a_0^0$ and $a_0^2$, we can use the whole data sample 
in one single fit, which will yield
the scattering length $a^0_0$, 
and the six form factor parameters 
$f_s$, $f_s^\prime$, 
$f_s^{\prime\prime}$, $g_p$, $g_p^\prime$, $h_p$.
The remaining form factor parameters  $f_e$, $\tilde{f}_p$, $g_e$, and 
$h_p^\prime$ have been fixed at zero. 
The results which are listed in Table~\ref{tab:ffwithuniversal}
are in excellent agreement with the ones derived in the
previous paragraph. However, as expected, the statistical errors
of the various parameters are smaller.
\begin{table}[htb]
\begin{center} 
\begin{ruledtabular}
\begin{tabular}{lc}
$f_s$                 & $ 5.75\pm0.02 \pm0.08$ (-0.03)\\  
$f_s^\prime$          & $ 1.06\pm0.10 \pm0.40$ (+0.37) \\ 
$f_s^{\prime\prime}$  & $-0.59\pm0.12 \pm0.40$ (-0.37) \\ 
 $g_p$                 & $ 4.66\pm0.05 \pm0.07$ (+0.03) \\ 
$g_p^\prime$          & $ 0.67\pm0.10 \pm0.04$  ($\pm$0.00)\\ 
$h_p$                 & $-2.95\pm0.19 \pm0.20$ (-0.16) \\ 
$a^0_0$ & $0.228\pm0.012\pm0.004$ 
($\pm$0.000) [Eq.~\ref{equ:universalcurve}]\\
$a^0_0$ & $0.216\pm0.013\pm0.004 $ ($\pm$0.000) 
[Eq.~\ref{equ:universalcurvegilberto}]\\
($\chi^2$/ndf& 30963/28793\\
\end{tabular} 
\end{ruledtabular} 
\end{center} 
\caption[Mass dependent fit]{Form factors and scattering length $a^0_0$
in the parameterisation of Eq.~\ref{equ:amoros} using either
Eq.~\ref{equ:universalcurve} or Eq.~\ref{equ:universalcurvegilberto}. 
The results for the form factor parameters are identical for both
fits. The first error is statistical, the second systematic.
The quantity in parentheses is
the shift in the result of the parameter which resulted from the
radiative corrections.   
\label{tab:ffwithuniversal}}
\end{table}
The quality of the fit can be judged from  Fig.~\ref{fig:mccompkin}.
The agreement between the Monte Carlo simulation modified for
the final values of the form factors and phase shifts
in all five kinematic variables is very satisfactory.
\begin{figure*}[htb]
\begin{center}
\includegraphics[width=105mm]{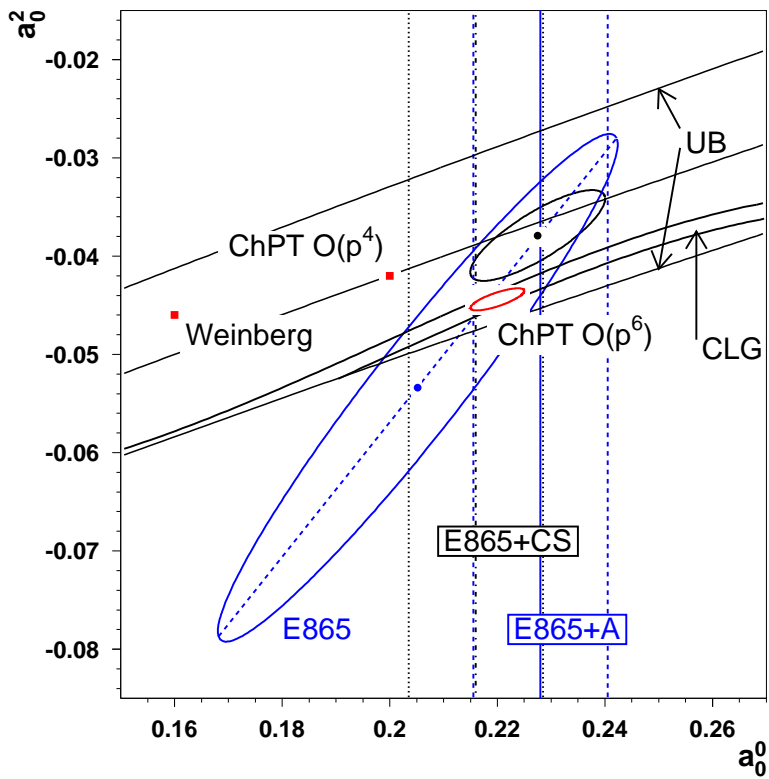}
\caption[Ellipse]{Results for the $\pi\pi$ scattering lengths
$a_0^0$ and $a_0^2$ obtained from fits to the $K_{e4}$ data directly or from
fits to the phase shifts obtained in this experiment.
Large ellipse labeled E865: 
fit to our $K_{e4}$-data leaving both $a_0^2$ and $a_0^0$ as free
parameters using Eq.~\ref{equ:schenk} with the parameters
of ref.~\cite{ananthanarayan00} (1$\sigma$ contour, see text for
remark concerning the region outside the universal band); medium size
ellipse without label: fit of ref.~\cite{descotes02} (1$\sigma$ contour)
to our phase shifts; theoretical predictions:
\cite{weinberg66} (Weinberg, square),
\cite{gasser83} (ChPT O(p$^4$, square), \cite{colangelo00} (ChPT O(p$^6$,
small ellipse);
solid curves labeled UB: universal band of allowed values
based on Eq.~\ref{equ:universalcurve}; solid curves labeled
CLG: narrow band of allowed values based on
Eq.~\ref{equ:universalcurvegilberto}; solid vertical line
labeled E865 (+A $\equiv$ analyticity constraints):
fit to $K_{e4}$ data using  Eq.~\ref{equ:universalcurve} with
1 $\sigma$ error limits given by dashed vertical lines;
dashed dotted line labeled E865 (+CS $\equiv$ analyticity and 
chiral symmetry constraints)
fit to $K_{e4}$ data using  
Eq.~\ref{equ:universalcurvegilberto} with
1 $\sigma$ error limits given by dotted vertical lines.
\label{fig:ellipse}}
\end{center}
\end{figure*}

In all previous fits, we have assumed that the decay rate does not depend on
$s_e$ and that there are no contributions from $p$-waves to $F$. To check 
this approximation we have allowed these form factors, one at a time, 
to vary in our fits too for the case where Eq.~\ref{equ:universalcurve} 
was used.
Table~\ref{tab:ffhigherorder} shows that all three form factors are found
to be consistent with zero. The nominal values of the
contributions to the form factors F and G are at the 2~\% or less
level. In all three cases, the dominant 
contribution to the systematic errors came from the resolution 
of the missing neutrino mass squared, and a smaller non-negligible  
error from the background estimate.
\begin{table}[htb] 
\begin{center}
\begin{ruledtabular} 
\begin{tabular}{lrc}
Parameters           & \multicolumn{1}{c}{Value}  & $\chi^2/\mbox{ndf}$ \\
$\tilde{f}_p$   & $ -0.34 \pm 0.10 \pm 0.27\, (- 0.02)$ & 30952/28792 \\ 
$f_e$           & $ -0.32 \pm 0.10 \pm 0.24\, (+ 0.02)$ & 30954/28792 \\ 
$g_e$           & $  0.04 \pm 0.34 \pm 0.88\, (\pm 0.00)$ & 30963/28792 \\ 
\end{tabular}  
\end{ruledtabular} 
\end{center} 
\caption[]{Results from the fits, where the form factors 
parameters 
$\tilde{f}_p$, $f_e$, and $g_e$ were allowed to vary one at a time.
The quantity in parentheses shows the influence of radiative corrections.
\label{tab:ffhigherorder}}
\end{table} 


In order to assess the sensitivity of our data to $a_0^2$ directly
we have also made a fit to the data where it was allowed to vary independently,
rather then being fixed via Eq.~\ref{equ:universalcurve} or
Eq.~\ref{equ:universalcurvegilberto}. The result is given in 
Table~\ref{tab:ffwouniversal} and Fig.~\ref{fig:ellipse}.
While the form factor parameters, as was expected, 
did not change, $a_0^0$ shifts to
a lower value with a larger error bar, which encompasses the
values found above. The error ellipse for this fit is shown in 
Fig.~\ref{fig:ellipse}. It illustrates the strong correlation between 
the two scattering lengths. The long axis of this ellipse follows
the equation $a_0^2=-0.1939+0.6851a_0^0$.  
\begin{table}[htb]
\begin{center}
\begin{ruledtabular} 
\begin{tabular}{lc}
$f_s$                 & $ 5.75\pm0.02 \pm0.08$ (-0.03) \\ 
$f_s^\prime$          & $ 1.06\pm0.10 \pm0.40$  (+0.37)\\ 
$f_s^{\prime\prime}$  & $-0.60\pm0.12 \pm0.40$  (-0.37)\\ 
$g_p$                 & $ 4.65\pm0.48 \pm0.07$  (+0.03)\\ 
$g_p^\prime$          & $ 0.69\pm0.11 \pm0.04$  ($\pm$ 0.00)\\ 
$h_p$                 & $-2.95\pm0.19 \pm0.20$  (-0.16)\\ 
$a^0_0$               & $0.203\pm0.033\pm0.004$ (-0.001)\\
$a^2_0$               & $-0.055\pm0.023\pm0.003$ (-0.001)\\
\hline 
$\chi^2/\mbox{ndf}$   & 30963/28792            \\ 
\end{tabular}  
\end{ruledtabular} 
\end{center} 
\caption[]{Fit of form factor parameterss and 
scattering lengths $a^0_0$ and $a^2_0$.
The first error is statistical, the second systematic. The quantity
in parentheses shows the influence of the radiative corrections. 
\label{tab:ffwouniversal}}
\end{table} 

\section{Summary and discussion}
\label{sec:conclusion}

The main results of this analysis are the measurements of the
$\pi\pi$-phase shift difference $\delta$ near threshold 
and of the form factors $F$, $G$ and $H$ of the hadronic current,
and their momentum dependence with a precision which has not
been previously attained. We emphasize again, that the analysis based on 
these data in six bins of invariant $\pi\pi$ mass is model independent.

The analysis which directly relates our data to the scattering 
length $a^0_0$, on the other hand, depends on
additional input, which leads to slightly different results.
While there is a consensus~\cite{colangelo01b,colangelo01a,descotes02}
on the use of the Roy-equations~\cite{ananthanarayan00} and 
Eq.~\ref{equ:schenk} to relate the phase shifts to the
scattering lengths, there exist slightly different ways of
linking $a_0^0$ to $a_0^2$, and how to make use
of peripheral $I=2$ data. These differences produce
slightly different results for both $a_0^2$ and $a_0^0$ with
overlapping statistical errors. The experimental 
and systematic uncertainities for both the phase shifts and scattering lengths
are considerably smaller than the statistical ones and are therefore
irrelevant to this discussion. 

If both $a_0^0$ and $a_0^2$ are allowed to vary independently
(Tab.~\ref{tab:ffwouniversal}), we obtain a result outside the
universal band in the  $(a_0^0,a_0^2)$ plane,
namely 
\[ a_0^0=0.203\pm0.033\ ,\;\; a_0^2=-0.055\pm0.023\ .\] 
Descotes {\it et al.}~\cite{descotes02}
have performed a fit to our published phase shifts~\cite{pislak01},
which are identical with the ones we give here
and obtained 
\[ a_0^0=0.237\pm0.033\ ,\;\; a_0^2=-0.0305\pm0.0226\ ,\]
with a strong correlation between the two values, which we also
observe in our result.
Only that
part of the 1$\sigma$ error contour of our result (the large ellipse in 
Fig.~\ref{fig:ellipse}) 
which overlaps  the universal band is consistent with both our and the $I=2$ 
data~\cite{hoogland74,losty74}, and only within this band the solution
of the Roy equations~\cite{ananthanarayan00} used here is 
valid~\cite{gilberto02}. From the 1$\sigma$ contour and 
its central axis we may deduce how much the results
listed in Tab.~\ref{tab:ffwithuniversal} change if the
input assumptions on the relation between $a_0^0$ and $a_0^2$ are varied.
Using  the lower
limit of the band defined by the bracket in Eq.~\ref{equ:universalcurve} 
we find a shift of $a_0^0$ by $-0.016$, while the maximum
allowed upward shift inside the $1\sigma$ contour and the
band is $0.012$. Assigning these values as theoretical
errors to our result, we obtain 
\begin{equation}\label{equ:a01} 
a_0^0=0.228\pm 0.012\;{\rm stat.}\;\pm0.004\;{\rm syst.}\;
^{+0.012}_{-0.016}\;{\rm theor.} 
\end{equation}
The use of Eq.~\ref{equ:universalcurve} implies
\begin{equation}\label{equ:a21} 
a_0^2=-0.0365\pm 0.023\;{\rm st.}\;\pm0.008\;{\rm sy.}\;
^{+0.0031}_{-0.0026}\;{\rm th.} 
\end{equation}
Since the central curve of the universal band is thought to be the best
representation of the $I=2$ data, it is no surprise, that the fit of
Descotes {\it et al.}~\cite{descotes02}, which used our
phase shifts and those of the Geneva-Saclay experiment~\cite{rosselet77}, 
Eq.~\ref{equ:schenk} with the parametrisation
of Ref.~\cite{ananthanarayan00} and the $I=2$ data below 800 
MeV~\cite{hoogland74,losty74}, gave nearly identical
results
\begin{equation}
a_0^0=0.228\pm 0.012\ ,\;\;
a_0^2=-0.0382\pm0.0038\ .
\label{equ:a022} 
\end{equation}
This result is also shown in Fig.~\ref{fig:scl}.

Using the narrower band in the $(a_0^2,a_0^0)$ plane defined by
Eq.~\ref{equ:universalcurvegilberto} our
result is
\begin{equation}\label{equ:a03}  
a_0^0=0.216\pm 0.013\;{\rm st.}\;\pm0.004\;{\rm sy.}\;
{\pm 0.002}\;{\rm th.}\ , 
\end{equation}
which implies
\begin{equation}\label{equ:a23} a_0^2=-0.0454\pm 0.0031\;{\rm st.}\;
\pm0.0010\;{\rm sy.}\;
{\pm 0.0008}\;{\rm th.}, \end{equation}
where the theoretical errors  have been evaluated as
before and correspond to the width of the band.
Descotes {\it et al.}~\cite{descotes02}, again fitting to our phase shifts, 
have obtained
for this case 
\begin{equation}\label{equ:a024}  
a_0^0=0.218\pm 0.013\ ,\;\; a_0^2=-0.0449\pm0.0033\ ,\end{equation}
again in agreement with our result and also with $a_0^0=0.221\pm 0.026$, 
obtained by Colangelo {\it et al.}~\cite{colangelo01a} by direct numerical
inversion of the relation between the phase shifts and the scattering lengths.

From this discussion we may deduce first that using our full data
sample or the phase shifts, which we have extracted from it, in the
six bins in $M_{\pi\pi}$ leads to the same results. This will make further use
of our data easy, should theoretical discussion continue and require this.
Second, it has become clear that the most probable values
of the two scattering lengths extracted from the
$K_{e4}$-data and low-energy $I=2$ data, resting on a minimum of
theroretical assumptions given by analyticity and 
crossing are those given in Eq.~\ref{equ:a01} and \ref{equ:a21}, or
Eq.~\ref{equ:a022}. 
Using the  additional constraints implied by chiral symmetry and
the value of the scalar radius~\cite{colangelo01a,colangelo01b}
leads to a value of the scattering length consistent within the
statistical errors with this result, albeit just 1$\sigma$ lower.
The authors of Ref.~\cite{descotes02} have elaborated in detail
how their ansatz differs from that of Ref.~\cite{colangelo01a},
and what the possible implications, if any, are for the
chiral pertubation theory parameters $\ell_3$ and $\ell_4$
and the size of the quark condensate. 
In view of the large errors and also inconsistencies
in the $I=2$ phase shift data~\cite{hoogland74,losty74}, it seems
premature to assign much significance to this minor discrepancy.
Because of the reduced theoretical
uncertainties we prefer to quote the values of
Eq.~\ref{equ:a03} and \ref{equ:a23} as our final
result. Both solutions for $a_0^0$ are in very good agreement with the
full two-loop standard ChPT prediction~\cite{colangelo00,colangelo01b}
\begin{center}\vspace*{-2mm}
$ a_0^0=0.220\pm 0.005\ ,\;\; a_0^2=-0.0444\pm0.0010\ .$
\end{center}

The influence of the reduced uncertainties of our
results on the form factors $F$, $G$ and $H$ on the
determination of the low 
energy constants of ChPT is evident from recent work of
Amor\'os {\it et al.}~\cite{talavera01}, who have updated their
earlier work~\cite{talavera00} using our data~\cite{pislak01}.
The constants $L_1^r$, $L_2^r$ and $L_3^r$ changed from
$0.53\pm0.25$, $0.71\pm0.27$ and $-2.72\pm1.12$ (in units of
$10^{-3}$), respectively, to $0.43\pm0.12$, $0.73\pm0.12$ and $-2.35\pm0.37$.

The first nonvanishing contribution to the anomalous form factor
$H$ in ChPT is predicted to be $H=-2.67$~\cite{wess71}. This agrees
well with our value of $H=-2.95\pm0.19\pm0.20$. An estimation of the
next to leading order gives only a small contribution~\cite{ametller93}.

\section*{Acknowledgements}

We gratefully acknowledge the contributions to the success of this 
experiment by Dave Phillips, the staff and management of the AGS
at the Brookhaven National Laboratory, and the technical staffs of the
participating institutions. We also thank J.~Bijnens, C.~Colangelo, J.~Gasser,
M.~Knecht, H.~Leutwyler, and J.~Stern for many fruitful discussions.
This work was supported in part by the U.S. Department of Energy,
the National Science Foundations of the U.S., Russia, and 
Switzerland, and the Research Corporation.


\begin{thebibliography}{99}
\bibitem[$^*$]{DB}{Now at: 
Rutgers University, Piscataway, NJ 08855}  
\bibitem[$^\ddagger$]{SE}{Now at: The Prediction Co., Santa Fe, NM 87505}
\bibitem[$^\S$]{HF}{Now at: Albert-Ludwigs-Universit\"at,
D-79104 Freiburg, Germany}
\bibitem[$^\|$]{JL}{Now at: University of Connecticut, Storrs, CT 06269}  
\bibitem[$^{**}$]{WMa}{Now at: LIGO/Caltech, Pasadena, CA 91125} 
\bibitem[$^{\ddagger\ddagger}$]{AS}{Now at:
        SCIPP, University of California, Santa Cruz, CA 95064}
\bibitem{shabalin63} E.P.~Shabalin, Sov. Phys. (JETP) {\bf 17}, 517 (1963)
(Zh. Eksp. Teor. Fiz. {\bf 44}, 765 (1963)). 
\bibitem{koller62} E.L. Koller {\it et al.}, 
Phys. Rev. Lett. {\bf 9}, 328 (1962).
\bibitem{okun60} L.B. Okun, and E.P.~Shabalin, Sov. Phys. (JETP) {\bf 10}, 
1252 (1960) (Zh. Eksp. Teor. Fiz. {\bf 37}, 1775 (1959)). 
\bibitem{birge65} R.W. Birge {\it et al.}, Phys. Rev. {\bf 139}, 1600 
(1965). 
\bibitem{ely69} R.P.~Ely~{\it et al.}, Phys. Rev. {\bf 180}, 1319 (1969).
\bibitem{schweinberger71} W.~Schweinberger~{\it et al.}, 
Phys. Lett. {\bf 36B}, 246 (1971).
\bibitem{bourquin71} M.~Bourquin {\it et al.}, Phys. Lett. {\bf 36B}, 615 
(1971); P.~Basile {\it et al.}, Phys. Lett. {\bf 36B}, 619 (1971); 
A.~Zylberstein {\it et al.}, Phys. Lett. {\bf 38B}, 457 (1972).
\bibitem{beier72} E.W.~Beier~{\it et al.}, Phys. Rev. Lett. {\bf 29}, 
         511 (1972); ibid. Phys. Rev. Lett. {\bf 30}, 399 (1973).
\bibitem{rosselet77} L.~Rosselet~{\it et al.}, Phys. Rev. D {\bf 15}, 
         574 (1977).
\bibitem{pislak01} S. Pislak {\it et al.}, 
Phys. Rev. Lett. {\bf 87}, 221801 (2001).
\bibitem{dirac} F. Gomez {\it et al.}, DIRAC Coll., Proc. Int. Euroconf.
on Quantum Chromodynamics: 15 Years of the QCD, Montpellier, France
(July 2000), Nucl. Phys. Proc. Suppl. {\bf 96}, 259 (2001).
\bibitem{ChPT} S.~Weinberg, Physica {\bf 96A}, 327 (1979); J.~Gasser 
        and H.~Leutwyler, Ann. Phys. {\bf 158}, 142 (1984);
        Nucl. Phys. {\bf B250}, 465 (1985).
\bibitem{leutwyler01} For a recent brief overview see e.g. H. Leutwyler, 
Proc. QCD@Work: Int. Conf. on QCD: Theory and Experiment, 
Martina Franca, Italy (June 2001), AIP Conf.Proc. {\bf 602}, 3 (2001);
hep-ph/0107332 
\bibitem{honnef02} 
{\em Effective field theories of QCD},
Proc. 264$^{\rm th}$ WE-Heraeus-Seminar, Bad Honnef, Germany (2001),
J. Bijnens, U.G. Mei\ss ner, and A. Wirzba eds.,  
hep-ph/201266, and references therein. 
\bibitem{gellmann68} M. Gell-Mann, R.J. Oakes and B. Renner, Phys. Rev. 
{\bf 175}, 2195 (1968).
\bibitem{GChPT} N.H.~Fuchs, H.~Sazdjian and J.~Stern, Phys. Lett. {\bf B269}, 
        183 (1991). 
\bibitem{knecht95} M. Knecht, B. Moussalam, J. Stern and N.H. Fuchs,
Nucl. Phys. {\bf B457}, 513 (1995); {\it ibid.} {\bf B471}, 445 (1996).
\bibitem{weinberg66} S.~Weinberg, Phys. Rev. Lett. {\bf 17}, 616 (1966). 
\bibitem{gasser83} J.~Gasser, and H.~Leutwyler, Phys. Lett. {\bf B125}, 
         325 (1983).
\bibitem{bijnens96} J.~Bijnens~{\it et al.}, Phys. Lett. {\bf B374}, 
         210 (1996); Nucl. Phys. {\bf B508}, 263 (1997); err. ibid.
         {\bf B517}, 639 (1998). 
\bibitem{colangelo00} C. Colangelo~{\it et al.}, Phys. Lett. {\bf B488}, 
         261 (2000). 
\bibitem{colangelo01b} G.~Colangelo, J.~Gasser, and H.~Leutwyler, 
Nucl. Phys. {\bf B603}, 125 (2001).
\bibitem{roy71} S.M.~Roy, Phys. Lett. {\bf 36B}, 353 (1971).
\bibitem{ananthanarayan00} B.~Ananthanarayan~{\it et al.}, Phys. Rep.
{\bf 353/4}, 207 (2001).
\bibitem{nagels79} M.M.~Nagels~{\it et al.}, Nucl. Phys. {\bf B147}, 
         189 (1979). 
\bibitem{talavera00} G. Amor\'os, J. Bijnens, and P. Talavera, Phys. Lett. 
{\bf B480}, 71 (2000); Nucl. Phys. {\bf B585}, 293 (2000); err. {\bf B598},
665 (2001)
\bibitem{talavera01} G. Amor\'os, J. Bijnens, and P. Talavera, Nucl. Phys.
{\bf B602}, 87 (2001).
\bibitem{ke494} For a recent discussion see J. Bijnens, C. Colangelo,
G. Ecker, and J. Gasser, {\em Semileptonic kaon decays}, hep-ph/9412392,
publ. in Ref.~\cite{daphne95}, p. 315.
\bibitem{daphne95} The Second DAPHNE Physics Handbook, L. Maiani, 
G. Pancheri, and N. Paver eds., INFN-LNF-Divisione Ricerca, SIS-Uffico 
Publicazioni, (Frascati 1995).
\bibitem{cabibbo65} N.~Cabibbo, and A.~Maksymowicz, Phys. Rev. {\bf 137},
         B438 (1965). 
\bibitem{pais67} A.~Pais, and S.B.~Treiman, Phys. Rev. {\bf 168}, 
         1858 (1968).
\bibitem{amoros99} G.~Amor\'os, and J.~Bijnens, J.~Phys. {\bf G25}, 
1607 (1999).
\bibitem{watson52} K.M.~Watson, Phys. Rev.{\bf 88}, 1163 (1952).
\bibitem{bijnens90} J.~Bijnens, Nucl. Phys. {\bf B337}, 635 (1990). 
\bibitem{riggenbach91} C.~Riggenbach~{\it et al.},  Phys. Rev. {\bf D43},
         127 (1991).
\bibitem{morgan94} For a review see D. Morgan, and M.R. Pennington, 
{\em Low energy $\pi\pi$ scattering}, publ. in
Ref.~\cite{daphne95}, p. 193.
\bibitem{morgan69} D.~Morgan, and G.~Shaw, Nucl. Phys. {\bf B10}, 261 (1969).
\bibitem{schenk91} A.~Schenk, Nucl. Phys. {\bf B363}, 97 (1991).
\bibitem{coeff} The numerical values for the coefficients are listed in
Ref.~\cite{ananthanarayan00}, appendix D, and Ref.~\cite{descotes02},
appendix B.
\bibitem{descotes02} S.~Descotes, N.H.~Fuchs, L. Girlanda, and J. Stern,
Eur. Phys. J. {\bf C24}, 469 (2002).
\bibitem{hyams73} B.~Hyams {\it et al.}, Nucl. Phys {\bf B64}, 134 (1973).
\bibitem{protopopescu73} S.D.~Protopopescu {\it et al.}, Phys. Rev. {\bf D7},
1279 (1973).
\bibitem{colangelo01a} G.~Colangelo, J.~Gasser, and H.~Leutwyler, 
Phys. Rev. Lett. {\bf 86}, 5008 (2001).
\bibitem{appel02} R.~Appel~{\it et al.}, Nucl. Instr. Meth. {\bf A479}, 349 
  (2002). 
\bibitem{atoyan92} G.S.~Atoyan~{\it et al.}, Nucl. Instr. Meth. {\bf A320},
         144 (1992).
\bibitem{groom00} D.E. Groom~{\it et al.}, Eur. Phys. J. {\bf C15}, 1 (2000);
K. Hagiwara {\it et al.}, Phys. Rev. {\bf D66} (2002) 010001.
\bibitem{diamant76} A.M.~Diamant-Berger, {\em \'Etude exp\'erimentale a haute
sta\-tis\-tique de la d\'es\-int\'e\-gration du meson $K^+$ dans le mode $K_{e4}$ et 
analyse des param\'etres qui gouvernent cette d\'es\-int\'e\-gration},
Ph.D. thesis, University of Paris, Orsay; Centre d'Etudes Nucl\'eaires 
de Saclay, 1976, report CEA-N-1918, unpublished.
\bibitem{ktaumat} see T.G.~Trippe, in Ref.~\cite{groom00}, p. 503.
\bibitem{kdalmat} K.O.~Mikaelian, and J.~Smith, Phys. Rev. {\bf D5}, 1763 
(1972). 
\bibitem{geant} R.~Brun~{\it et al.}, GEANT 3.21, CERN Geneva.
\bibitem{appel99} R.~Appel~{\it et al.}, Phys. Rev. Lett. {\bf 83}, 4482
(1999); H.~Ma~{\it et al.}, Phys. Rev. Lett. {\bf 84}, 2580
(2000).
\bibitem{eadie71} W.T.~Eadie~{\it et al.}, {\it Statistical Methods in
Experimental Physics}, (North Holland, Amsterdam and London, 1971). 
\bibitem{lafferty95} G.D.~Lafferty and T.~R.~Wyatt, Nucl. Instr. Meth.
{\bf A355}, 541 (1995).
\bibitem{diamant76a} see Ref.~\cite{diamant76}, appendix 2.
\bibitem{neveu68} A. Neveu, and J. Scherk, Phys. Lett. {\bf 27B}, 384 (1968).
\bibitem{alavi01} A. Alavi-Harati {\it et al.}, Phys. Rev. {\bf D64}, 112004
(2001). 
\bibitem{low58} F. Low, Phys. Rev. {\bf 110}, 974 (1958).
\bibitem{hoogland74} W. Hooogland {\it et al.}, Nucl. Phys. {\bf B69}, 266
  (1974); ibid. {\bf B126}, 109 (1977).
\bibitem{losty74} M.J. Losty {\it et al.}, Nucl. Phys. {\bf B69}, 185 (1974).
\bibitem{gilberto02} We thank G. Colangelo for pointing this out.
\bibitem{wess71} J.~Wess and B.~Zumino, Phys.~Lett. {\bf 37B}, 95 (1971).
\bibitem{ametller93} L.~Ametller~{\it et al.}, Phys.~Lett. {\bf B303}, 140 
        (1993).
\end{thebibliography}
\end{document}